\documentclass{aa}
\usepackage{amssymb,amsfonts,amsmath,times,pict2e}
\usepackage{natbib}
\usepackage{graphicx}
\usepackage{enumerate}
\usepackage{subfigure}
\usepackage{bm}
\usepackage{hyperref}

\usepackage{color}
\usepackage{ulem}
\definecolor{mygray}{gray}{0.6}
\definecolor{magenta}{rgb}{0.858, 0.188, 0.478}

\newcommand{\Rmnum}[1]{\expandafter\@slowromancap\romannumeral #1}

\usepackage{xspace}
\newcommand{\fg}[1]{Fig.~\ref{fig:#1}}
\newcommand{\eq}[1]{Eq.~(\ref{eq:#1})}
\newcommand{\Eq}[1]{Equation~(\ref{eq:#1})}%beginning of the sentence
\newcommand{\se}[1]{Sect.~\ref{sec:#1}}
\newcommand{\Se}[1]{Section~\ref{sec:#1}}%beginning of the sentence
\newcommand{\noeq}[2]{$#1$\strut$#2$}
\newcommand{\tb}[1]{Table~\ref{tab:#1}}%beginning of the sentence

\newcommand{\AU}{ \  \rm AU}
\newcommand{\K}{ \  \rm K}
\newcommand{\Ms}{ \  \rm M_\odot}
\newcommand{\Rs}{ \  \rm R_\odot}
\newcommand{\yr}{ \   \rm yr}

\newcommand{\g}{ \   \rm g}
\newcommand{\cm}{ \   \rm cm}

\newcommand{\kG}{ \ \rm kG}
\newcommand{\Me}{ \ \rm M_\oplus}

\newcommand{\taud}{ \tau_{\rm d}}
\newcommand{\tauc}{ \tau_{\rm c}}
\newcommand{\taum}{ \tau_{\rm m}}

\usepackage{CJKutf8}

\begin{document}
\begin{CJK*}{UTF8}{gbsn}

\title{Dynamical rearrangement of super-Earths during disk dispersal}
%\title{Rebound of super-Earths during gas-disk dispersal}
\subtitle{I. Outline of the magnetospheric rebound model}

%\author{ Beibei Liu(刘倍贝) \inst{1}, Chris W. Ormel\inst{1}, Douglas N.C. Lin(林潮)\inst{2,3,4,5} }
\author{ Beibei Liu \inst{1}, Chris W. Ormel\inst{1}, Douglas N.C. Lin\inst{2,3,4,5} }

\institute{Anton Pannekoek Institute (API), University of Amsterdam, Science Park 904,1090GE Amsterdam, The Netherlands\label{inst1}
\and Department of Astronomy and Astrophysics, University of California, Santa Cruz, CA 95064, USA\label{inst2}
\and Institute for Advanced Studies, Tsinghua University, Beijing, 100086, China\label{inst3}
\and Kavli Institute for Astronomy \& Astrophysics, Peking University, Beijing, 100871, China\label{inst4}
\and National Astronomical Observatory of China, Beijing, 100012, China\label{inst5}\\
\email{[b.liu@uva.nl,c.w.ormel@uva.nl,lin@ucolick.org]}
   }

\date{\today}

\abstract{The Kepler mission has discovered  that close-in super-Earth planets are common around solar type stars. They are often seen together in multiplanetary systems, but their period ratios do not  show strong pile-ups near mean motion resonances (MMRs). One scenario is that super-Earths form early, in the presence of a gas-rich disk. Such planets interact gravitationally with the disk gas, inducing their orbital migration.  Disk migration theory predicts, however, that planets would end up at resonant orbits due to their differential migration speed. 
}{Motivated by the discrepancy between observation and theory, we seek for a mechanism that moves planets out of resonances. We examine the orbital evolution of planet pairs near the magnetospheric cavity during the gas disk dispersal phase.  Our study  determines the conditions under which planets can escape  resonances.}{We extend Type I migration theory by calculating the torque a planet experiences at the interface of  the empty magnetospheric cavity and the disk: the one-sided torque. We perform two-planet N-body simulations with the new Type~I expressions, varying the planet masses, stellar magnetic field strengths, disk accretion rates and gas disk depletion timescales.}{As planets migrate outward with the expanding  magnetospheric cavity, their dynamical configurations can be rearranged.  Migration of planets is substantial (minor) in a massive (light)  disk.  When the outer planet is more massive than the inner planet, the period ratio of two planets increases through outward migration. On the other hand, when the inner planet is more massive, the final period ratio tends to remain similar to the initial one.  Larger stellar magnetic field strengths result in planets stopping their migration at longer periods.  We  apply this model to two  systems, Kepler-170 and Kepler-180. By fitting their present dynamical architectures, the disk and stellar B-field parameters at the time of disk dispersal can be retrieved. } {We highlight \textit{Magnetospheric rebound} as an important ingredient able to reconcile disk migration theory with observations.  Even when planets are trapped into MMR during the early gas-rich stage, subsequent cavity expansion would induce substantial changes to their orbits, moving them out of resonance.} 

\maketitle

\section{Introduction}

Together the \textit{Kepler} and \textit{K2} transit missions have discovered over $3\,300$ exoplanets and $500$ multi-planet systems. These data  constitute a large  sample to statistically analyze  the properties of exoplanets.   The majority of planets discovered by Kepler are close-in super-Earths -- planets with radii $R_{\rm p}<4 R_{\oplus}$, or mass $M_{\rm p}\lesssim 10 M_{\oplus}$, and  periods $P <100$ days.  Super-Earths are very common: nearly half of solar-type stars harbor a super-Earth(s) \citep{Petigura2013}.  This occurrence rate appears to be independent (or at most weakly dependent) on stellar type and metallicity \citep{Howard2012,Fressin2013,Mayor2011,Bonfils2013,Mulders2015}. Several tens of super-Earths have their masses known from follow-up radial velocity or transit timing variation \citep{Lithwick2012b,Marcy2014}. Combined with the radii, we are able to obtain the bulk densities. For many of them, bulk densities are too low to be consistent with a pure rocky composition: a hydrogen-helium atmosphere is required. In order to acquire such a substantial gaseous atmosphere,  super-Earths are inferred to have formed in early the gas-rich disk phase, before the depletion of the disk gas \citep{Lopez2014,Rogers2015}.

 Super-Earths are frequently found in multiple, compact systems with relatively low eccentricities and inclinations \citep{Johansen2012,Fang2012,Fabrycky2014,Shabram2016,Xie2016}.
 Their orbits neither exhibit strong pile-ups at mean motion resonances (MMRs) nor are their period ratios uniformly distributed (see Fig.~6 of \cite{Winn2015}). In particular, the period ratio distribution shows an asymmetry around major resonances with a deficit abundance just interior to and an excess slightly exterior to the $2$:$1$ and $3$:$2$ MMRs.  Also, a significant fraction of planets are  found with period ratios much larger than $2$. 

Planets embedded in disks gravitationally interact and transfer angular momentum with the disk gas, resulting in their orbital migration. For low mass planets, this mechanism is known as  Type I migration \citep{Lin1979,Goldreich1979,Kley2012,Baruteau2014}. Theoretically, resonance capture  is a natural outcome of planet migration  \citep{Lee2002,Papaloizou2005,Pierens2008}.   The existence of some resonant systems, like Kepler-223 \citep{Mills2016}, proves the fidelity of disk migration theory. However, as mentioned above, statistically, the majority of super-Earths are not in MMRs.  This discrepancy is a key mismatch between observation and theory.  

 Different scenarios  have been proposed to explain the observed deviation from exact resonance.  These include tidal damping of planets \citep{Lithwick2012,Delisle2012,Batygin2013,Lee2013,Xie2014,Delisle2014}, planet mass growth \citep{Petrovich2013}, interaction with planetesimals \citep{Chatterjee2015}, stochastic torques in  turbulent disks \citep{Rein2012,Batygin2017}, planet-wake interaction \citep{Baruteau2013}, and resonant overstability \citep{Goldreich2014,Delisle2015}.  But  these models are mostly limited to small departures -- at a few percent level -- from resonance.  Other scenarios \citep{Ogihara2009,Cossou2014,Ogihara2015}  propose that giant impacts scatter planets away  from original resonances after the dispersal of the disk gas.  However, these models may not be able to explain the  above asymmetry around major resonances.  The onset of this orbital instability also requires a sufficient amount of embryos (typically $N$$\gg$$2$ )  in  a compact configuration at early stage.  Nevertheless,  neither of above scenarios have considered  the influence of the stellar magnetic field.

T Tauri stars are magnetically active with observed field strengths of  kilogauss-level at their surfaces  \citep{Johns-Krull2007}. For comparison, the \textit{current} B-field of the Sun is only $0.3 \rm \ G$. The stellar magnetic field will truncate the disk at the magnetospheric cavity radius \citep{Koenigl1991}. This radius is determined  by  equating  the stellar magnetic torque with the viscous torque of the disk.  Adopting  parameters  for a typical T Tauri star,  the magnetospheric cavity radius is around $0.1 \AU$.  The size of this cavity increases with time as the viscous torque diminishes during the  disk dispersal.   In this paper we explore how the expanding cavity affects the dynamical evolution of super-Earth planets.

We propose a new model -- \textit{magnetospheric rebound}  -- which considers the migration of planets near the magnetospheric cavity.
Generally, during the disk dispersal, planets -- trapped into MMR -- can migrate outward together with the expanding magnetospheric cavity. However, when the cavity expansion rate is high or when the disk is no longer massive, the planet will be left behind. Therefore, each planet subsequently drops into the cavity at the time determined by  their mass, the disk mass and the gas depletion timescale. As a consequence, planets can substantially alter their dynamical configuration with the cavity expansion during gas disk dispersal, in some cases resulting in the break-up of the MMR.
%\ccc{the next few sentences are too detailed; you don't want to introduce things as 'cavity expansion timescale'. Also unclear yet for the reader why we focus on 2 planets. Try to stay general.} For only one planet and at the proximity of the disk edge,  the net torque it experience is positive. So the planet is able to keep migrating outward with the expanding cavity until  its migration timescale exceeds the cavity expansion timescale. When two planets are trapped into resonance near the cavity, they rebound with each other in resonance  and migrate outward  together with the cavity expansion. This resonance is distrupted at the point where the inner planet drops into the cavity but  the outer planet migrates outward.   Their divergent migration leads to the resonant escaping. Therefore, planets can substantially alter their dynamical configuration with the cavity expansion during gas disk dispersal \ccc{this is a nice sentence}. 

 In this paper, we will test this  \textit{magnetospheric rebound}  model by N-body simulations. The key parameters, apart from the planet masses, are the initial disk accretion rate, the gas disk depletion timescale, and the stellar magnetic field strength. In \se{method}, we start by presenting the disk model and review the Type I migration torque formulas.  In \se{results}, we carry out numerical simulations and  demonstrate  the illustrated runs. Two observed Kepler systems are modelled in our parameter study in \se{obs}. 
Finally, we  discuss the results and draw conclusions  in \se{conclusion}.

\section{Method}
\label{sec:method}
In this section,  we give a description about the adopted disk model (\se{disk}) including the magnetospheric cavity, which truncates the disk at a radius $r_{\rm c}$ (\se{cavity}), and provide  expressions for the Type I torques that act on the planets  (\se{TypeI}). \Se{applicability} outlines under which conditions the model is applicable.

\subsection{Disk model}
\label{sec:disk}
We consider a disk model with a constant aspect ratio 
\begin{equation}
h = \frac{H}{r} = 0.025,
\label{h} 
\end{equation} 
where $H $ is the disk scale height and $r$ is the disk radius.  The corresponding gas disk temperature is: 
\begin{equation}
T_g =  \left( \frac{ \mu h^2 G M_\ast}{r R_{\rm g}}  \right)= 150  \left( \frac{r}{1  \AU} \right)^{-1} \rm \ K,
\label{eq:temp}
\end{equation}
where $G$ is gravitational constant, $R_{\rm g}$ is gas constant, $M_\ast$ is the stellar mass, and $\mu$ is the molecular weight in the protoplanetary disk.  
The gas surface density ($\Sigma$) is derived from the gas accretion rate $\dot{M}_g$ by the steady-state assumption for viscous disks ($\dot M_{g} = 3\pi \Sigma \nu$). We use the  \cite{Shakura1973} $\alpha$ prescription  for viscosity, 
$ \nu  = \alpha_{\nu} H^2 \Omega $, where $\Omega$ is the Keplerian angular frequency. We focus on the migration of close-in super-Earths  where the disk region  is primarily MRI turbulent. Therefore, $\alpha_{\nu} =  10^{-2}$  is  adopted in this paper.  The gas surface density  is then given by 

\begin{equation}
%\Sigma =  24 \left(  \frac{\dot M_g}{10^{-9} \Ms \yr^{-1} }\right) \left(  \frac{\alpha_{\nu}}{10^{-2} }\right)^{-1}  \left(  \frac{r}{1 \AU}\right)^{-0.5}  \g \cm^{-2}.
\Sigma =  75 \left(  \frac{\dot M_g}{10^{-9} \Ms \yr^{-1} }\right) \left(  \frac{\alpha_{\nu}}{10^{-2} }\right)^{-1}  \left(  \frac{r}{0.1 \AU}\right)^{-0.5}  \g \cm^{-2}.
\label{eq:gasdensity}
\end{equation}

Our study concerns the late phase of disk evolution when the gas disk dispersal takes place.  We assume  the disk gas accretion rate follows: 
 \begin{equation}
{\dot M}_{\rm g} =  {\dot M}_{\rm g0}   \exp\left[- t/\taud\right],
\label{eq:taudep}
\end{equation} 
where $\tau_d$ is the characteristic disk depletion timescale in late stage. We note that $\tau_d$ is shorter than the typical lifetime of the disk ($\sim2$-$3~\rm Myr$, \cite{Mamajek2009}) by at least one order of magnitude \citep{Williams2011}.

  %$t$ is the time that the disk starts its rapid decay. 
  
\subsection{Inner magnetospheric cavity}
\label{sec:cavity}

Young T Tauri stars are observed with strong magnetic fields of \noeq{\approx}{0.2}--6 kG \citep{Johns-Krull2007,Yang2011,Johns-Krull2013}. The stellar magnetic field lines are strongly coupled to  disk  gas  close to the central star.    The Lorentz torque is generated by the stellar-disk magnetic interaction while the viscous torque is induced by the disk gas.   The inner disk is truncated at the radius where the Lorentz torque is greater than the viscous torque \citep{Ghosh1979a,Koenigl1991,Armitage2010}. The gas flow near the cavity edge is accreted  onto the stellar surface along field lines, which is known as magnetospheric accretion. We assume that the gas inside the cavity ($r<r_{\rm c}$) is removed very quickly, and therefore that the inner edge of the disk is sharply cut-off (see \fg{1side}).   The magnetic torque per unit area is $B^2 r/2\pi$ and the viscous torque is $\dot {M_{\rm g}} r^2 \Omega_{\rm K}$.  Assuming that the central star has a dipole magnetic field ($B(r) = B_{\ast} R_{\ast}^3/r^3$) aligned with the stellar rotation axis, the size of this magnetospheric cavity reads \citep{Frank1992,Armitage2010}:
\begin{equation}
 r_c = \left(  \frac{ B_{\ast}^4 R_\ast^{12}  }{  4GM_\ast \dot M_{g}^2}  \right)^{1/7},
 \label{eq:r_cavity}
\end{equation}
where $G$ is the gravitational constant,  $B_{\ast}$ is the magnetic field strength at the stellar surface, $R_\ast$ is the stellar  radius and  $\Omega_{\rm K}$ is the angular velocity at distance $r$.
Adopted from the fiducial value of T Tauri stars ($M_\ast=1 \Ms$, $R_\ast= 2 \Rs$), \eq{r_cavity} evaluates as
\begin{equation}
  \begin{split}
r_{\rm c} = 0.1  \left(  \frac{ \dot M_{g}}{10^{-9}   \Ms  \yr^{-1}}\right) ^{-2/7} \left(  \frac{ B_{\ast} }{ 1  \kG}\right)^{4/7 }  \AU.
%r_{\rm c} =  &0.1  \left(  \frac{ \dot M_{g}}{10^{-9}   \Ms  \yr^{-1}}\right) ^{-2/7} \left(  \frac{ B_{\ast} }{ 1  \kG}\right)^{4/7} \\
% &  \left(  \frac{ M_{\ast}}{\rm 1 \Ms }\right)^{-1/7} \left(  \frac{ R_{\ast} }{ 2 \ R_{\oplus}}\right)^{12/7}    \AU.
\label{eq:r_c}
\end{split}
\end{equation}
From \eq{r_c},  it is clear that  the  cavity  expands with decreasing $\dot M_{g}$, a consequence of the diminishing importance of the viscous torque.

\Eq{r_c} is based on two assumptions: (i) the stellar magnetic field is dipole; and (ii) the magnetic axis is aligned with the stellar rotation axis.  In reality, the field can contain higher-order contributions, for example quadrupole term ($B_{\ast} R_{\ast}^4/r^4$), which could dominate at stellar distances $r\sim R_\ast$. 
%In that case the magnetic torque is an even more strongly decreasing function of radius. 
However, at larger distances -- important for this study -- the dipole contribution will take over. 
%For the same field strength at the stellar surface and disk accretion rate, the truncation radius $r_{\rm c}$ is closer to the central star in a quadrupole filed than dipole field.
Also, both theoretical analyses \citep{Lipunov1980,Lai2011} and MHD simulations of magnetospheric accretion \citep{Romanova2003,Romanova2004} indicate that when the magnetic field axis is misaligned with the rotation axis of the disk, the inner disk is likely to be warped. In that case the gas density distribution is not axisymmetric as in \eq{gasdensity}. Nonetheless, many of the features that we use in the axisymmetric case would still be present: there would still be a magnetospheric cavity, which expands during the disk dispersal, and there would still be a one-sided planet-disk interaction (as discussed in Sect. 2.3). 
%These are the key ingredients for the magnetic rebound model. 
However, to model this additional complexity is beyond the scope of this paper.
%Clearly, this requires a more complex model, which is beyond the scope of this paper. Nevertheless, we do }
 
 \begin{figure}[t]
    \centering
         \includegraphics[scale=0.4, angle=0]{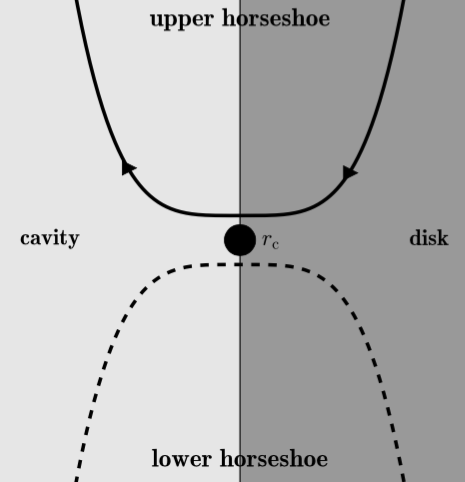}
       \caption{  Sketch of the  gas motion near the cavity edge. The gas only completes the upper horseshoe U-turn before it is accreted to the central star by magneto-stellar forces, which we assume to remove the gas very quickly. There is no gas  left to execute the lower horseshoe.
    }
    \label{fig:1side}
\end{figure} 
 
With decreasing $\dot M_{\rm g}$ the truncation radius ($r_{\rm c}$) will become larger than the corotation radius $r_{\rm co}$ -- the radius where the disk Keplerian frequency  equals the spin frequency of the star. In principle, when $r_{\rm c}$$>$$r_{\rm co}$, accretion is quenched and the angular momentum is transferred from the stellar spin to the disk.  However, this process also leads to a stellar spin down and the expansion of the corotation radius. Provided the disk depletion time is sufficiently long or that the spin synchronization proceeds rapidly, $r_{\rm c}$ and the corotation radius expand in tandem and accretion  onto the host star is maintained. Observationally, Kepler target stars have modest spin periods up to a few months \citep{McQuillan2014}, generally longer than the orbital periods of the inner-most super-Earths.  Therefore, these systems could have experienced an expansion of the corotation radius at the time of disk dispersal.    In this work we will, for simplicity, assume that the disk truncation radius is  given by \Eq{r_c}  and that the disk always accretes onto the star.

\subsection{Type I torques acting on the planet}
\label{sec:TypeI}
Having specified a model for the gas structure, we now discuss the backreaction of the disk on the planet. The key point in our discussion is the distinction between two-sided torques and one-sided torques. A sketch illustrating the difference between one-sided and two-sided torques is given in \fg{torque}.   The purple and blue arrows denote the Lindblad and corotation torques, respectively. Two-sided torques apply when the planet is far away from the disk cavity (\fg{torque}a). In that case it experiences the inner and outer Lindblad torques from both sides. Usually, the net (differential) value is negative, causing inward migration \citep{Ward1997}. The corotation torque is determined by the gradient of disk surface density and temperature across the horseshoe region of the planet. It is described in \se{2-sided}.   One-sided torques are applicable when the planet is at the cavity edge  (\fg{torque}b). In that case it  experiences a negative, one-sided Lindblad torque and a positive one-sided corotation torque. One-sided torques  are generally much larger than two-side torques, because the the near-cancellation effect -- a feature of the two-sided torques -- is absent. The descriptions of one-sided Lindblad torque  and one-sided corotation torque will be given in \se{1-sided Lindblad} and  \se{1-sided corotation}.  In \se{combine} a general Type I torque  that combines these two regimes  is presented.  

\begin{figure}[t]
    \includegraphics[scale=0.29, angle=0]{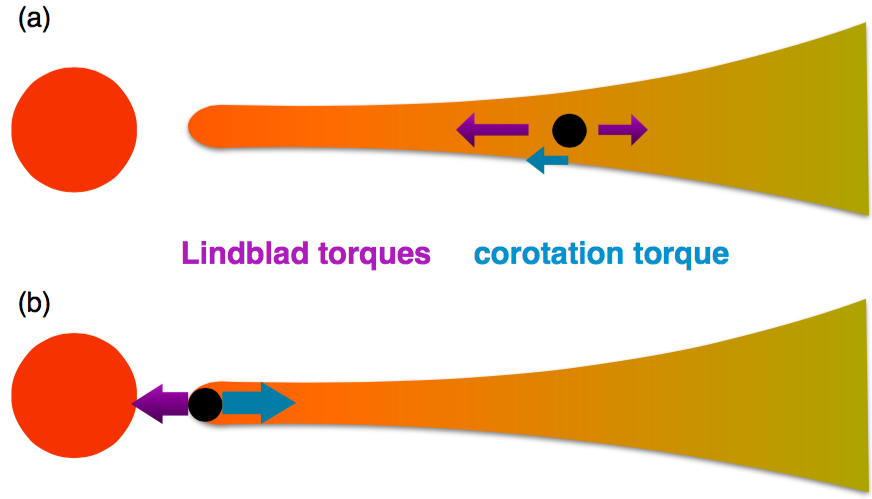}
    \caption{
       Sketch illustrating the difference between two-sided and one-sided torques. (a) When the planet is far away from the disk cavity, it experiences  two sided-torques from the  disk. (b) When the planet is at the cavity edge,  only one-sided torques operate. The purple and blue arrows denote the Lindblad and corotation torques, respectively. 
        %In the two-sided limit, the planet  feels the inner and outer Lindblad torques from its both sides, and the differential value is generally inward; the corotation torque is determined by the gradient of disk surface density and temperature across the horseshoe region of the planet. When the planet is at the cavity edge (bottom), it only experiences an negative, one-sided Lindblad torque and a positive one-sided corotation torque.
        }
    \label{fig:torque}
\end{figure} 

\subsubsection{Two-sided torques}
\label{sec:2-sided}
 When the planet is far from the disk edge and embedded in a continuous disk,  the interior disk exerts a positive torque and pushes the planet outward, whereas the exterior disk pushes it inwards.  To first order, these contributions cancel. The `standard'  two-side Lindblad torque ($\Gamma_\mathrm{L,2s}$) is therefore a differential torque  with the sign being determined by the gradient of the disk structure \citep{Goldreich1980,Tanaka2002}.  In addition, the planet interacts with the gas within the co-orbital horseshoe region (the two-sided corotation torque).  

Considering these two torque components, we adopt the total two-sided torque from Eq. (49) of \cite{Paardekooper2010} (local isothermal approximation),
\begin{equation}
\begin{split}
    & \frac{\Gamma_\mathrm{2s}} {m_{\rm p} (r_p \Omega_p)^2}  =    \frac{\Gamma_\mathrm{L,2s} + \Gamma_\mathrm{c,2s} } {m_{\rm p} (r_p \Omega_p)^2}\\
     &=  \left[-(2.5 + 0.5\beta +0.1s) +  1.4\beta + 1.1(\frac{3}{2} +s) \right] q_{d} \frac{q_p}{h^2},
    \label{eq:2s}
    \end{split}
\end{equation}
where $\beta$ and $s$ are the gradient of temperature and gas surface density, $q_d \equiv \Sigma_p r_p^2 / M_{\ast}$ and $q_p\equiv m_p/M_{\ast}$ are dimensionless measures of the local disk mass and planet mass, respectively.  The notation $X_p$ indicate quantity $X$ evaluated at the  planet location $r = r_p$.  The first term on right side of \eq{2s} describes the two-sided Lindbald torque ($\Gamma_{\rm L,2s}$) and other two terms represent the two-sided corotation torque ($\Gamma_{\rm c,2s}$).

\subsubsection{One-sided Lindblad torque}
\label{sec:1-sided Lindblad}
When a planet approaches the inner edge of the disk, however, the torques from the  disk exterior to planet start to dominate over those from  the interior disk. In the limit of a vanishing interior disk, the planet only experiences  negative Lindblad torques from the exterior disk.  No longer is the torque a result from a near-cancellation of two large contributions. Instead, the torque has become a first order effect. We denote  this torque as $\Gamma_\mathrm{L,1s}$.

The one-sided Lindblad torque can be calculated from the impulse approximation \citep{Lin1979, Lin1993}: 
\begin{equation}
    \Gamma_\mathrm{L,1s}  
    %= \int_{\Delta_0}^{\infty}  2 \pi \Sigma_p r_p j \ {\rm d}\Delta 
=-  \frac{8(Gm_p)^2 r_p \Sigma_p}{9 \Omega_p^2}   \int_{b_{\rm min}}^{\infty}\frac{ {\rm d} b}{b ^4},
\end{equation}
where 
%$j$ is the specific angular momentum transfer rate, 
$b=r-r_p$ is the separation between each annular gas parcel  and the planet and $b_{\rm min}= 2H/3$ is a cutoff boundary adopted for the torque density \citep{Ward1997,Artymowicz1993}.  We then obtain for the one-sided Lindblad torque:
\begin{equation}
    \frac{\Gamma_\mathrm{L,1s}} {m_{\rm p} (r_p \Omega_p)^2} = C_{L} q_{d} \frac{q_p}{h^3},
    \label{eq:lindblad1}
\end{equation}
where the prefactor $C_{L}  = -0.65$.
%$C_{L} = 2^5/3^5 \left[   2K_0(2/3) +K_1(2/3) \right ]^2 \simeq 0.84$

\subsubsection{One-sided corotation torque}
\label{sec:1-sided corotation}
We next calculate the corotation torque when the planet is at the cavity edge. In the context of a co-rotation torque with one-sided, it  means that only the upper horseshoe is present (where material is transported from orbits exterior to the planet to orbits interior to the planet; see \fg{1side}). The lower horseshoe motion is absent  by virtue of the assumption that any gas interior to the planet is quickly removed along magnetic field lines and accreted to the central star.
Therefore, the planet exchanges angular momentum only through the upper horseshoe, which provides a strong positive corotation torque. We denote this torque the one-sided corotation torque $\Gamma_\mathrm{c,1s}$. It can be calculated from angular momentum conservation principles, i.e., the angular momentum lost due to material being pushed to a lower Keplerian orbit is gained by the planet. Following \citet{Paardekooper2009a}: 
 \begin{equation}
    \Gamma_\mathrm{c,1s} = \int_{0}^{x_{\rm hs}} \Sigma (j-j_p) (\Omega - \Omega_p) r \ {\rm d}r =  \frac{1}{2} (\Omega_p r_p)^2 \Sigma r_p^2 x_{\rm hs}^{3}, 
 \label{eq:hs_torq}
\end{equation}
where  $j=\Omega_K(r)r^2$ is the specific angular momentum corresponding to $r$.  We adopt $x_{\rm hs} =1.7 (q_p/h)^{0.5}r_p$  \citep{Paardekooper2009b, Ormel2013} for the half-width of the horseshoe region and obtain
 \begin{equation}
    \frac{\Gamma_\mathrm{c,1s}} {m_{\rm p} (r_p \Omega_p)^2} = C_\mathrm{hs} q_{d}  \left(\frac{q_p}{h^{3}}\right)^{1/2}, 
\label{eq:hs1}
\end{equation}
where $C_\mathrm{hs} =  2.46$.

Based on the assumption that gas is removed quickly at the edge of the disk, the surface density has a infinite sharp transition at $r_{\rm c}$. Under this assumption, the planet  obtains its maximum positive one-sided corotation torque and it  maximizes the rebound (outward migration). For simplicity, we only consider this situation. Instead, if gas removal is not an efficient  process, there could be a more gradual transition, with $\Sigma$ being (locally) a power-law. In that case, the Lindblad and corotation torques at $r_{\rm c}$ are  determined by the two-sided torques expression, which reduces (but does not diminish) the rebound. Three-dimensional MHD simulations of magnetospheric accretion \citep{Romanova2002} nevertheless  reveal a strong cut-off of the gas density near the disk edge, supporting our approximation.

%We  note that  $\Gamma_\mathrm{2s}$  (both for $\Gamma_\mathrm{L,2s}$ and  $\Gamma_\mathrm{c,2s}$)  is a second order torque, which sign depends on the gradient of the disk profile in the vicinity of the planet. But $\Gamma_\mathrm{2s}$ is  independent of $s$ and $\beta$.  $\Gamma_{\rm L,1s}$ is always negative and $\Gamma_{\rm c,1s}$ is always positive.  Based on Eqs. (\ref{eq:2s}), (\ref{eq:lindblad1}) and (\ref{eq:hs1}) and by order of magnitude estimation,  $|\Gamma_{L,1s}|\simeq |\Gamma_{L,2s}|/h$, and $|\Gamma_{\rm c,1s}| \simeq |\Gamma_{\rm c,2s}|/h$ when $q_p \simeq h^{-3}$  (appropriate for super-Earths at $ 0.1 \AU$).   

 The migration rate corresponding to a torque ($\Gamma$) is 
\begin{equation}
    \frac{\dot a}{a} = \frac{ 2\Gamma}{ m_{\rm p} r_p^2 \Omega_p}.
\label{eq:adot}
\end{equation}
In \fg{mig}, the migration rate ($\dot a/a$) is plotted for the two-sided torque $\Gamma_{\rm 2s}$, the one-sided corotation torque $\Gamma_{\rm c,1s}$, the one-sided Lindblad torque $\Gamma_{\rm L,1s}$ as  function of $q_p$ at $r=0.1 \AU$. The dark grey zone refers to the  non-linear regime  and the light grey zone indicates the gap opening regime (see the discussion in  \se{applicability}).
%The slopes of the curves  presented  in \fg{mig} are different because of  the planet mass dependence on $\Gamma$.  The Lindblad  torque is a  backreaction from the disk density perturbation, so it scales with $m_{\rm p}^2$.   However, the corotation torque increases with the horseshoe width such that  $\Gamma_{\rm c,1s} \propto  x_{\rm hs}^{3}$ (\eq{hs_torq}) and $\Gamma_{\rm c,2s} \propto  x_{\rm hs}^{4}$ ($x_{\rm hs} \propto m_{\rm p}^{1/2}$). Thus,  the migration rate  of one-side corotation goes slower than that of one-side Lindblad  and two-side torques.    
Two-sided torques are second order torques, which sign depends on the gradient of the disk profile in the vicinity of the planet (\eq{2s}). But one-sided torques are  independent of $s$ and $\beta$ (\eq{hs1} and \eq{lindblad1}). Indeed, one-sided torques are larger than two-sided torques (approximately by a factor $h^{-1}$).  Between the one-sided torques, the positive corotation torque is larger than the negative Lindblad torque when $q_p$ is small. 
Recall that  the one-sided torques operate for the disk region near the cavity whereas the two-sided torques operate for the disk region away from the cavity. Therefore, small planets are able to migrate outward from the disk edge until the point where the two-sided torques dominate. This also means that small planets would migrate outward when the cavity gradually expands.
    
However, when the planet is large enough to enter the gap-opening regime, the corotation torque diminishes due to the depletion of gas in the horseshoe region.  Then, the (negative) one-sided Lindblad torque would be in magnitude  larger than the (positive) one-sided corotation torque. In that case, the planet directly migrates into the cavity and is left behind as the disk cavity ($r_c$) expands.
\begin{figure}[t]
    \includegraphics[width=88mm, angle=0]{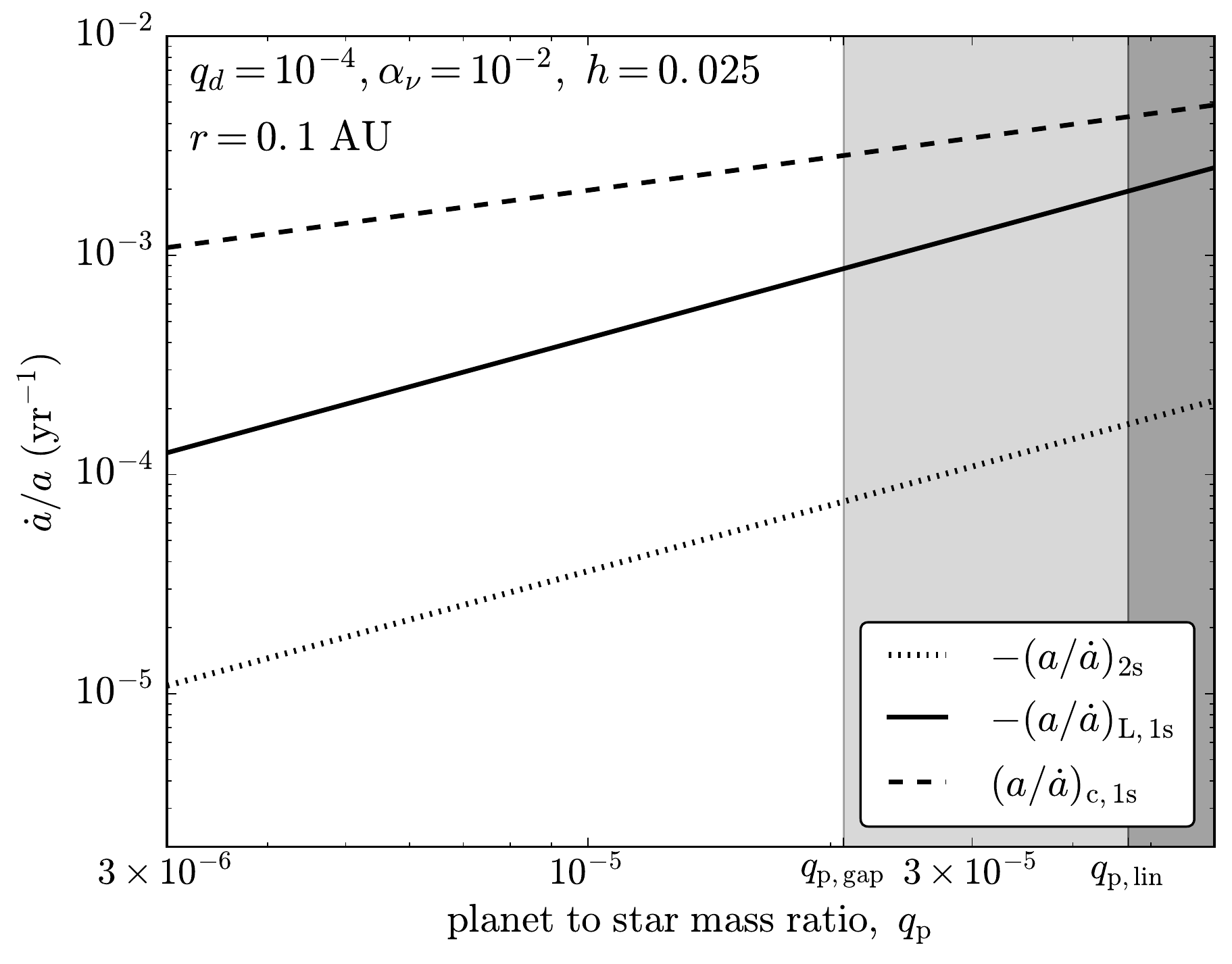}
    \caption{
        Planet migration rate ($\dot a/a$) for different torques as a function of dimensionless planet mass $q_p=M_p/M_\star$ ar $r=0.1 \AU$. 
        The dotted, solid and dashed lines correspond to the  migration rate estimated from \eq{adot} when   $\Gamma = \Gamma_{\rm 2s}$ (two-sided torque), $\Gamma_{\rm L,1s}$ (one-sided Lindbldd torque) and $\Gamma_{\rm c,1s}$ (one-sided corotation torque), respectively. The dark grey zone refers to the  non-linear regime, and the light grey zone indicates  the gap opening regime.  $q_{\rm p,gap}$ and $q_{\rm p,lin}$ are the critical masses for these two regimes (see \se{applicability}).  The adopted disk parameters are $q_{\rm d} =10^{-4}$, $\alpha_{\nu} =10^{-2}$ and  $h =0.025$.   
        }
    \label{fig:mig}
\end{figure}

Since the eccentricities of planets could be excited by  planet-planet interaction,  we also consider  the saturation of the corotation torque (both one-sided and two-sided) \citep{Bitsch2010}:
    \begin{equation}
    \Gamma_{\rm c}(e)  =  \Gamma_{\rm c} (0)  \exp\left[-e/e_{\rm f}\right],
\label{eq:sat}
\end{equation}
where $e$ is the eccentricity of the planet, $\Gamma_{\rm c} (0)$ is the corotation torque for zero eccentricity, and $e_{\rm f} = h/2 +0.01$ \citep{Fendyke2014}.

\subsubsection{Combined Type I torque}
\label{sec:combine}
A more general case appiles when a planet approaches the disk edge, which falls in between the one-sided and embedded (two-sided) regimes discussed above.  In that case, the torque expression is approximated by interpolating  the one-sided and two-sided torques:
 \begin{equation}
\Gamma =
 \begin{cases}
 f \Gamma_{1s} + (1-f) \Gamma_{2s} & \mbox{when $r \geq r_c$} \\
   0 & \mbox{when $r<r_c$}
 \end{cases}
\end{equation}
where the coefficient $f =  \exp \left[-(r-r_c )/x_{\rm hs}\right]$ is a measure for the proximity of a planet to the disk edge ($f=1$: at the edge; $f=0$: far away from the edge).  The form of the expression ensures that  the total torque is dominated  by the one-sided  torque  when the planet is located within a half-horseshoe width from the disk edge ($r_p \leq r_c + x_{\rm hs}$).

%Note the one-sided torque we obtain here based on  infinite sharp disk edge.  If disk edge has a finite width, the new one-sided torque will be a fraction of  the one-sided torque we derived  in above shape edge case, $\Gamma_{1s,fw} = f \Gamma_{1s}$ where magnitude of  f ($0 < f \leq 1$) represents this (sharp or gradual) transition of the disk edge.  

\subsection{Model applicability}
\label{sec:applicability}
Our expressions for the Type-I torques are valid  in the linear regime where the planet's perturbation is small. This requires the planet Hill radius to be smaller  than the disk scale height   ($R_{H}  \equiv (m_p/3M_{\ast})^{1/3}  <H$, \cite{Lin1993}). This condition can be expressed as
\begin{equation}
    q_p \lesssim 3h^3 = 4.7\times10^{-5} \left( \frac{h}{0.025} \right)^3
  \label{eq:q_linear}
\end{equation}
The no-gap opening criterion also requires that the planet torque ($\Gamma_p$)  is smaller than the viscous torque ($\Gamma_{\nu} =3 \pi \Sigma \nu r^2 \Omega$). Otherwise the disk does not supply enough material (angular momentum) to fuel the planet.   This condition is especially relevant to planets at the disk edge $r_c$ where the planet torque is dominated by the 1-sided corotation torque, $\Gamma_p= \Gamma_{\rm c,1s}$ (\eq{hs1}).  The no-gap opening condition reads
\begin{equation}
\left(\frac{q_p}{h^3}\right)^{3/2} \lesssim \frac{3 \pi \alpha_{\nu}}{C_{\rm hs} h},
\label{eq:q_nogap}
\end{equation}
or
\begin{equation}
    q_p \lesssim 2.1\times10^{-5} \left( \frac{h}{0.025} \right)^{7/3} \left( \frac{\alpha_\nu}{10^{-2}} \right)^{2/3}.
\end{equation}

    For typical super-Earth planets ($M_{p} \lesssim 10 \ M_{\oplus}$) around solar type stars,  the above two conditions  require $h \gtrsim 0.025$ near the edge of the disk. In addition, the no-gap opening condition requires a large $\alpha_\nu$ parameter. This explains our default values for $h$ and $\alpha_\nu$.   The $\Gamma_{\rm c,1s}$ is the torque that provides planet outward migration nearly the cavity edge. For very low $h$ and $\alpha_\nu$ disks,   gaps can be opened by super-Earth planets.   $\Gamma_{\rm c}$ is reduced since the gas is severely depleted in their horseshoe regions.  As a consequence,  we do not expect the magnetospheric rebound mechanism to operate in systems with low $h$ and $\alpha_\nu$.  

    But how realistic are our adopted values of $h$ and $\alpha_\nu$? Typically, MHD simulations modelling the MRI \citep{Balbus1998} indicate $\alpha \simeq 10^{-2}$ \citep{Davis2010,Suzuki2010}, provided the disk is sufficiently ionized.
  Generally, this holds true for the very inner disk ($\gtrsim 10^3 \rm \ K$) where thermal ionization of alkali elements produce the required amount of electrons \citep{Armitage2011}.  As the cavity radius ($r_{\rm c}$)  expands, however, the temperature drops and ionization  becomes dependent on non-thermal processes (\textit{e.g.}, stellar X-ray irradiation). Although initially the high densities should prevent  X-rays  from penetrating  the disk midplane, the expansion of the cavity radius also quickly decreases $\Sigma$ since  $\Sigma(r_c)\propto r_c^{-4}$ (see Eqs. (\ref{eq:gasdensity}) and (\ref{eq:r_cavity})). In addition, we note that because we study the inner disk ($a\lesssim 0.3 \AU$) at late times, small dust grains have coagulated into large bodies. In such a dust-free medium the liberated electrons always end up in the gas, increasing the electron fraction and promoting the MRI. 
 
The disk's aspect ratio is  determined by a thermal balance: an increased midplane temperature increases $h$ (\eq{temp}).  
    A high midplane temperature can result from viscous dissipation (a high $\dot{M}$) in combination with a large vertical optical depth. But as the disk evolves at late stage, a higher $T$ can also arise when the (visible) optically depth is low and stellar photons penetrate the disk deeper.
    % Since we are looking at the stage where late disk dispersal phase, this situation can be expected. 
    For example, in the optically thin limit, the equilibrium temperature at 0.1 AU for a solar mass star would be $1200 \K$, corresponding to a scale height of 0.022.  In addition,  magnetic accretion generates  additional heat  near the cavity edge \citep{Hartmann1998}, further increasing the aspect ratio $h$.
    
    Based on these considerations, we believe that the adopted values of $h =0.025$ and $\alpha=10^{-2}$ are reasonable. Note that  we assume a constant aspect ratio in this paper, which is independent of $r$. Another aspect ratio profile, $h = 0.05 r^{1/5}$,  has also been investigated and we find that the results presented in the following sections are only weakly affected. Overall, our results should be applicable to super-Earth planets as long as the inner disk is indeed hot enough.

  In addition, the magnetospheric accretion process requires efficient coupling between the field lines and the disk gas. This condition implies that the magnetic diffusivity $\eta$  is sufficiently small, $\eta \lesssim H c_{\rm s}$  \citep{Wardle2007}.   
    The criterion for the onset of MRI is that the magnetic Reynolds number is larger than $1$, $Re_{\rm M} = v_{\rm A} H/\eta = \alpha_{\nu} ^{1/2} H c_{\rm s}/\eta \gtrsim 1$ \citep{Armitage2010}, where $c_{\rm s}$ and $v_{\rm A}$ are the sound speed and  Alfv\'en speed.  This suggests $\eta \lesssim  \alpha_{\nu} ^{1/2} H c_{\rm s} $.
    Therefore, the disk  always satisfies the coupling condition once the MRI is operational.  The model is applicable since we study super Earth systems with typical orbital periods of a few weeks around their host stars. 
     
 %The required non-negligible ionization fraction can be sustained in the inner region where alkali metals are thermally ionized ($\gtrsim 10^3 \rm \ K$). 
 %   As the cavity radius ($r_{\rm t}$) expands further out where the temperature and gas density decreases, the non-thermal ionization processes take the role (\textit{ e.g.}, stellar X-ray irradiation,  radioactive decay of $^{26}$Al). Because we study the inner disk region ($a\lesssim 0.3 \AU$) and in late disk depletion phase, small dust grains are effectively coagulate into large bodies. In a dust-free case the liberated electrons end up in the gas, increasing the electron fraction.
 %  On the other hand, photons can penetrate deep into the disk midplane due to the low gas column density with the cavity expansion ($\Sigma\propto r_c^{-4}$, see Eqs. \ref{eq:gasdensity} and \ref{eq:r_cavity}).  Both two factors increase  the ionisation fraction substantially 

\cite{Ogihara2010} also calculate a torque near the disk edge, which they term the `edge torque'. This torque operates for eccentric planets at the disk edge, which experience asymmetric eccentricity damping from the disk gas.  The magnitude of the edge torque increases with the eccentricity of the planet.   However, when calculating the total torque in their Eq.~(17), they still use the  two-sided Type I torque expression. As we demonstrate in this paper, when the planet is near the disk edge, one-sided torques will dominate over two-sided torques.  Also, we find that the eccentricities of the  planets are modest (a few times $10^{-2}$) in our simulations. For these low eccentricities, the one-sided torque (\eq{hs1}) is much larger than the edge torque (Eq. (15) of \cite{Ogihara2010}).  We therefore neglect the edge torque in this paper.

\section{Results}
\label{sec:results} 

\begin{table*}
    \centering
    \caption{Adopted parameters}
    \begin{tabular}{lllllllllllll|}
        \hline
        \hline
        Run  & Name & $M_\mathrm{in}$   & $M_\mathrm{out}$ & $\dot M_{g0}$   & $\tau_{\rm d}$   & $B_{\ast}$  & \multicolumn{2}{c}{Period ratios}  \\
                &   &   ($M_{\oplus}$) &($M_{\oplus}$)    & ($ \rm M_{\odot} yr^{-1}$)  &  ($ \rm yr$)  &  ($\kG$)  & Initial  & Final \\
        \hline
       \#1 & Default        & 3 & 6 & $10^{-8}   $& $10^5$  & 1 & 1.75& 3.24  \\
       \#2 & Massive inner  & 6 & 3 & $10^{-8}   $& $10^5$ & 1  &  1.75 & 1.57  \\
       \#3 & Strong $\rm B_\ast$       & 3 & 6 & $10^{-8}  $& $10^5$ & 5 &1.75 &  3.23   \\
       \#4 & Light disk     & 3 & 6 & $5 \times 10^{-10}  $& $10^4$ & 1 & 1.75 & 1.72     \\
    \hline
        \hline
    \end{tabular}
    \label{tab:tab1}
\end{table*}

\begin{figure*}[t]
    \includegraphics[scale=0.5]{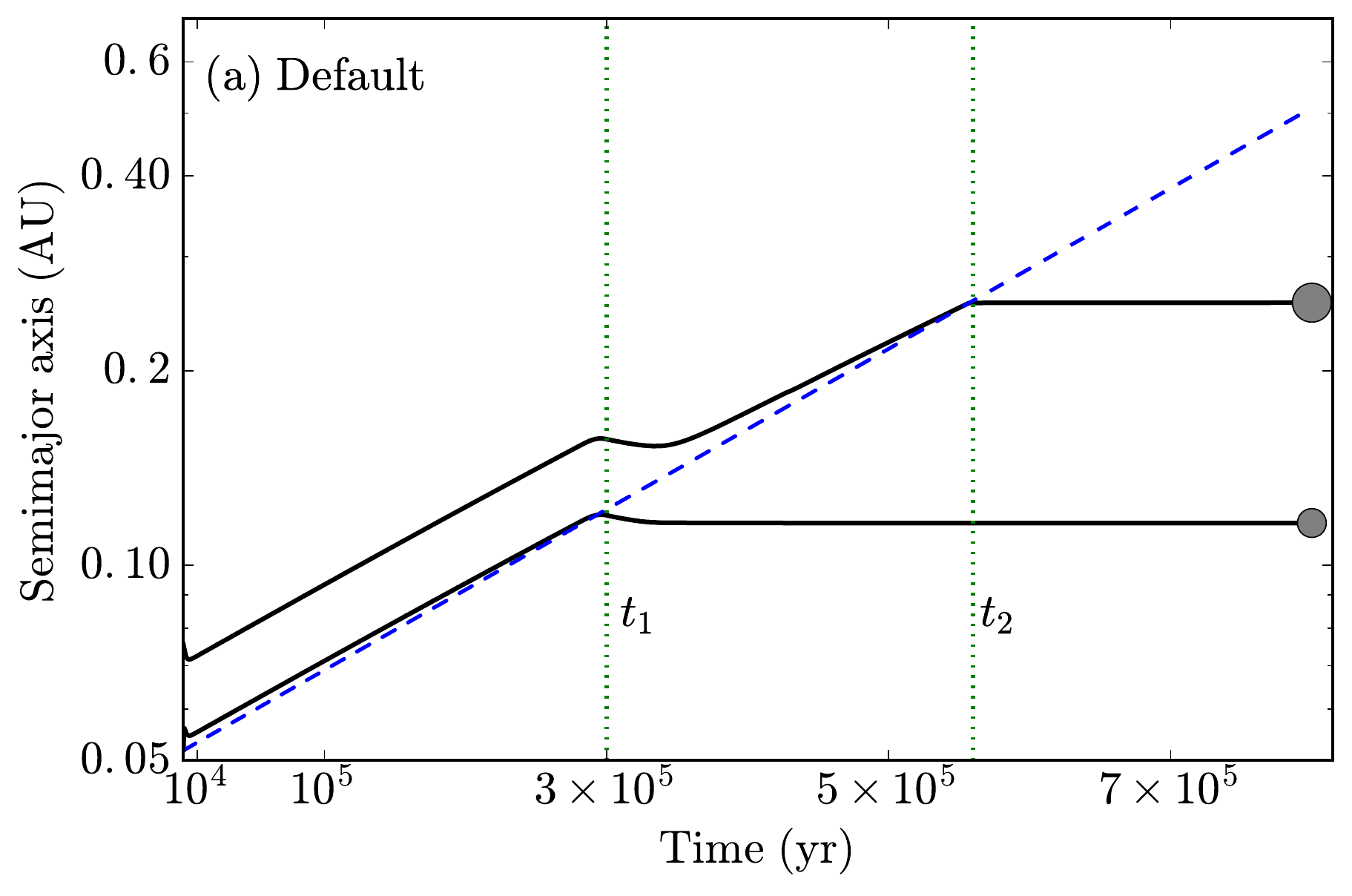}
    \includegraphics[scale=0.5]{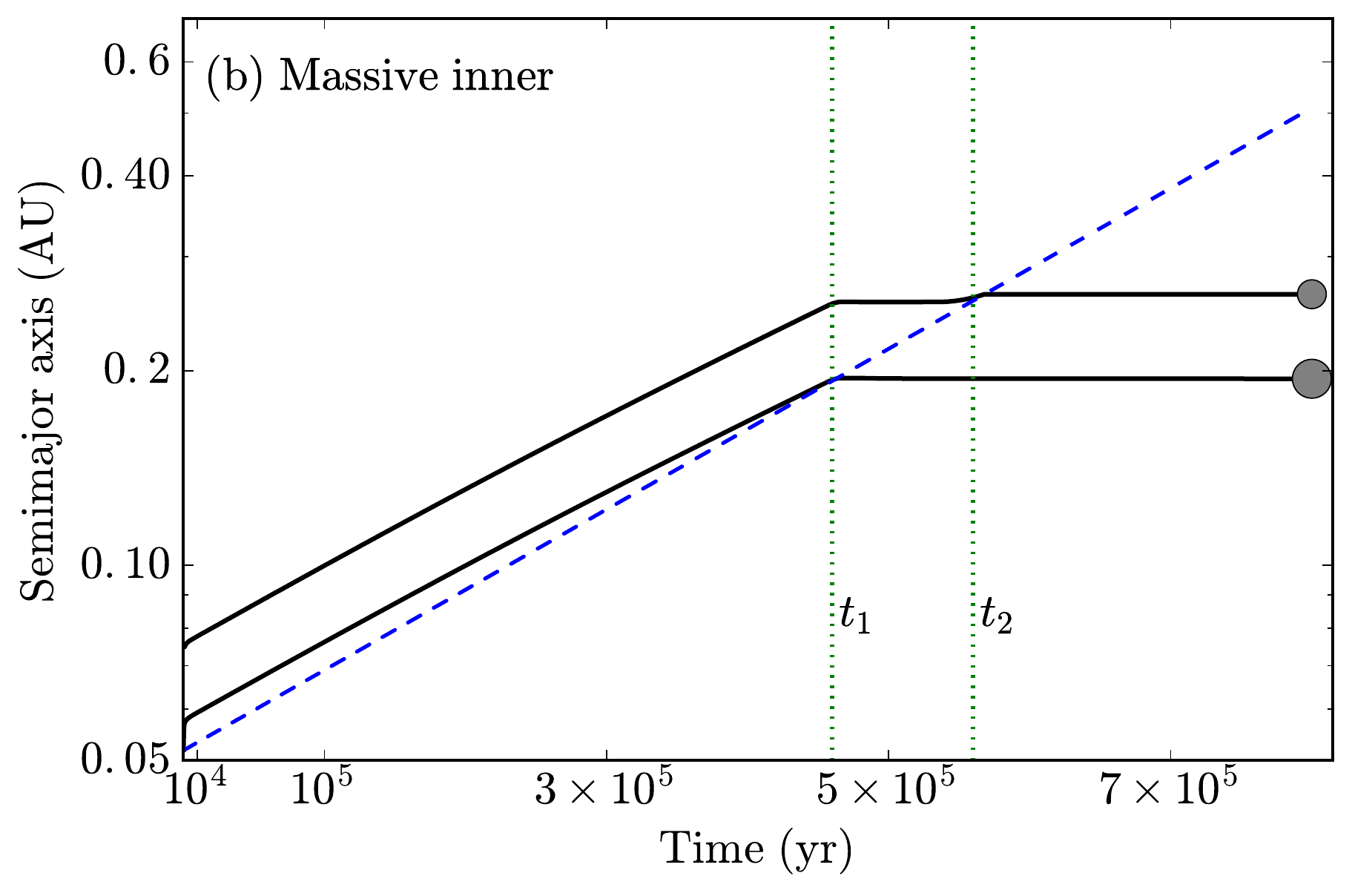}
    \includegraphics[scale=0.5]{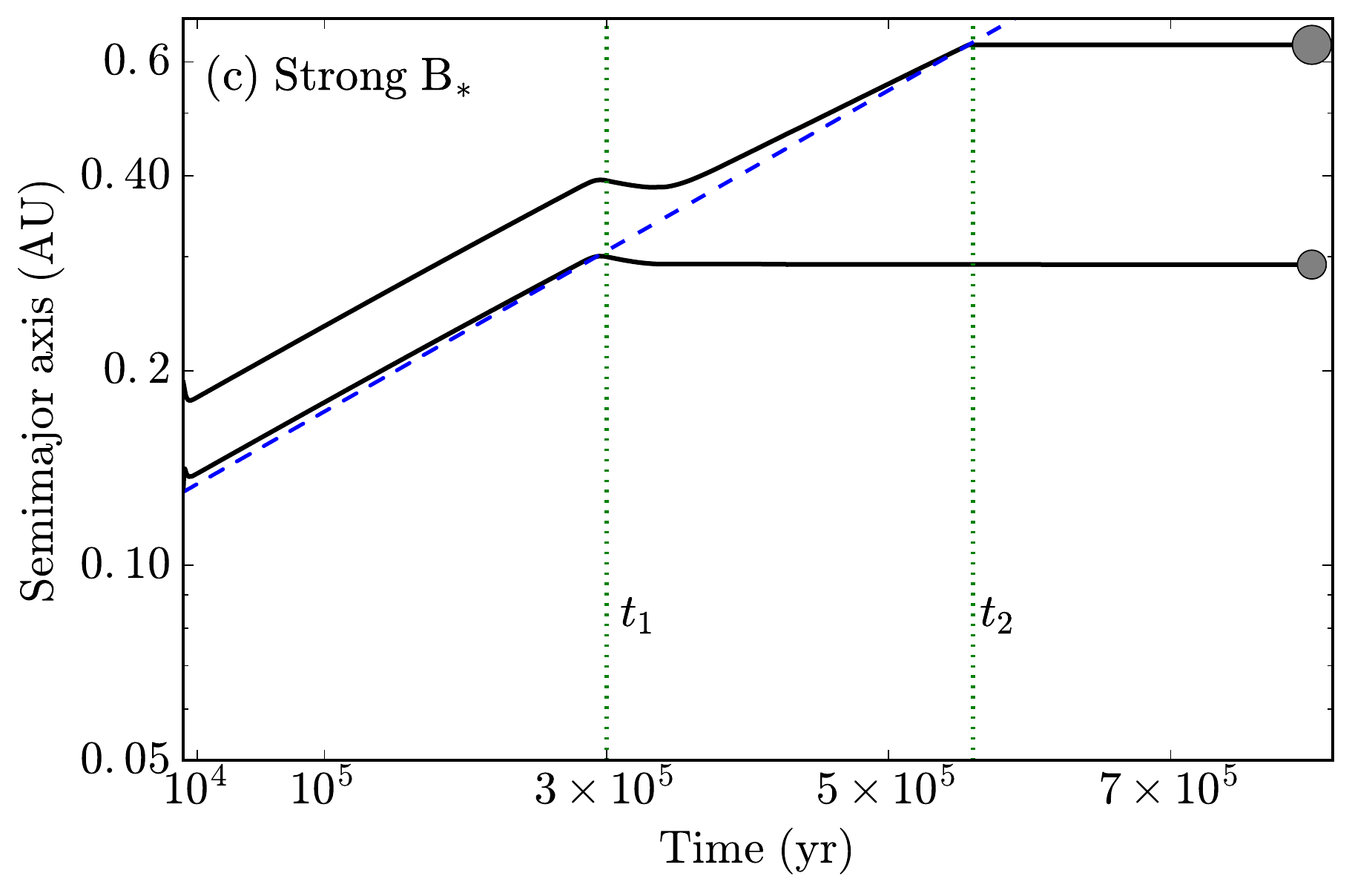}
    \includegraphics[scale=0.5]{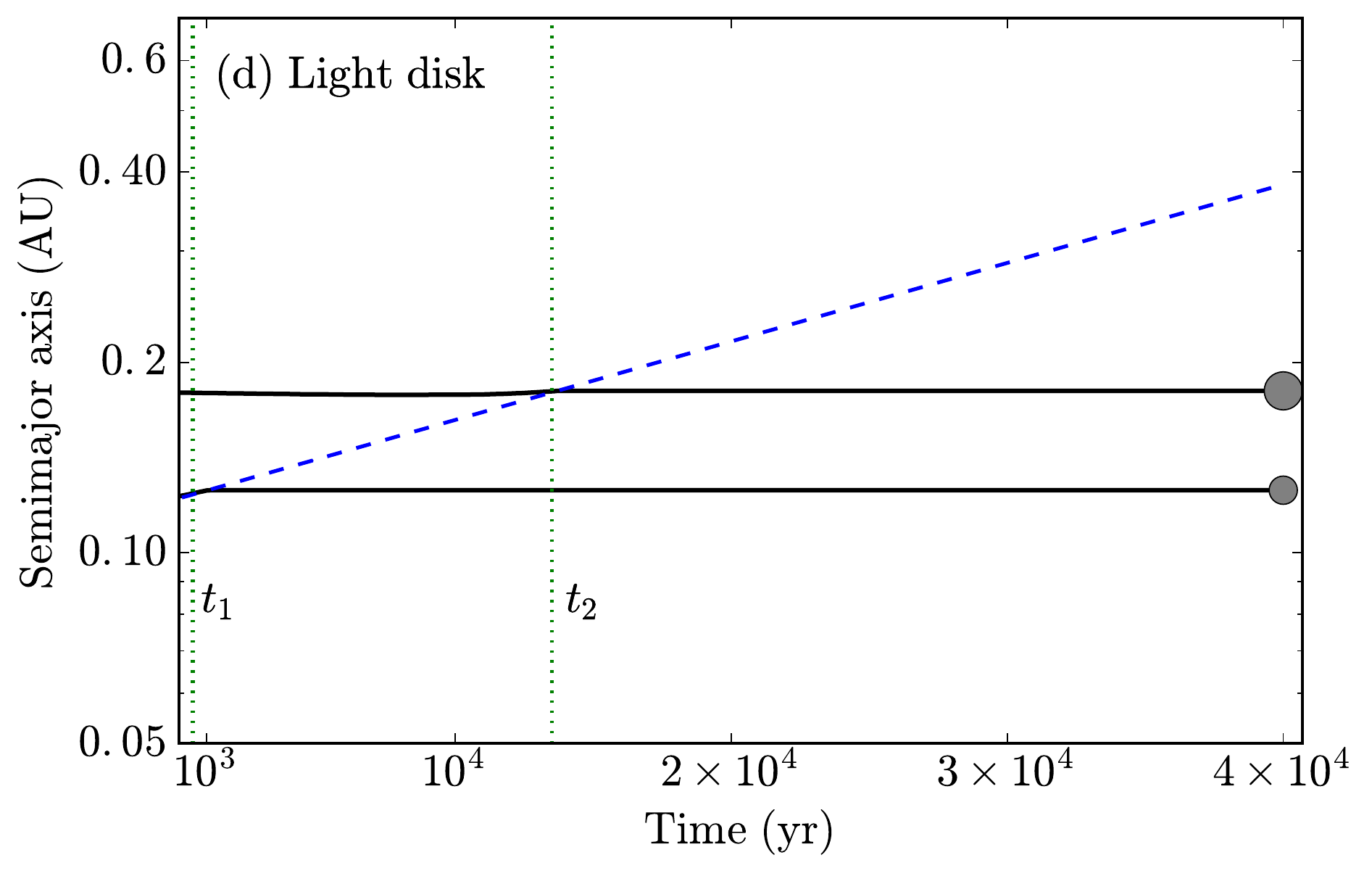}
    \caption{Migration of a planet pair in the  gaseous disk.  Black lines trace  the time evolution of the semi-major axes of the planets and the blue dashed line represents the inner magnetospheric radius $r_c$. The four panels show simulations with  (a) default parameters, (b) massive inner planet, (c) strong stellar magnetic field and (d)  a light  disk . Parameters are listed in Table 1. The vertical lines represent the time when the inner ($t_1$) and the outer planet ($t_2$) enter the cavity.  Note that  the the scale of the x-axis of panel (d) is different from the others.} 
\label{fig:fig2}
\end{figure*}

 %Added the Type I torque formulas into our Hermite-Embryo N-body code\ccc{you write this as if everybody is familiar with your code.} (the detail of the code is introduced in   \cite{Liu2015}), we are able to carry out the numerical experiments\ccc{too long sentence with 2 ideas: (i) N-body code with added torques; (ii) differences to Liu+2015}. 

We carry out numerical experiments on Hermite N-body code \citep{Aarseth2003}. In this code, the planet-planet interaction is calculated with Hermite scheme, and the planet-disk interaction is  included by adding additional torque recipes (See details in \cite{Liu2015}).  However,  \cite{Liu2015} only consider the two-sided torques \citep{Paardekooper2011}.   In this work, we add the new torque formulas demonstrated in above section,  which is a combination of both one-sided and two-sided torques.  A further difference is that this work considers  in the  locally isothermal limit, which is applicable to the  late disk dispersal  phase when the disk opacity is substantially reduced. 

The goal of our simulations is to investigate the evolution of the period ratio of the planet pair during the dispersal of the gas disk. In this paper, only two-planet systems are considered.  The simulations are performed until the disk gas is entirely depleted.  Because the long-term secular evolution of a two-planet system would not change their orbits once the planets are in  the Hill stable regime, we can consider the orbits of planets at the end of simulations as their present-day orbits.  
 A set of simulations (referred to as run 1--4) with different planet masses ($M_\mathrm{in}$ and $M_\mathrm{out}$), disk properties ($\dot{M}$, $\tau_d$), and stellar magnetic field strength ($B_\star$) are performed.  We set  $ M_\ast = 1 \ \rm M_\odot$ throughout this section.  The adopted parameters are given in Table \ref{tab:tab1}  and the results are  illustrated in \fg{fig2}.

\subsection{Planet pair with massive outer one} 
The result from the default run is presented  in \fg{fig2}a. The  parameters are set as  $ \dot M_{\rm g0}= 10^{-8} \Ms \yr^{-1} $, $\taud=10^{5}  \yr$ and $B_{\ast}= 1\kG$.  The black solid lines show the time evolution of  embryos' semimajor axes and the blue dashed line represents  the expansion of  $r_{\rm c}$.  Initially, the magnetospheric cavity radius ($r_{\rm c}$) is at  $0.052 \AU$  and  two embryos are located at $0.052 \AU$ and  $0.076 \AU$. Their orbits are hence initialized between the $3$:$2$ and  $2$:$1$ MMR.  The inner and outer planet masses are $3 \Me$ and $ 6 \Me$.

Because of convergent migration, the planets soon get trapped into the $3$:$2$ MMR and migrate inward to the magnetospheric cavity within a few thousand years.  When the inner planet gets close to the cavity edge ($r_{\rm in} \simeq r_{\rm c}$),  one-sided torques become dominant. For this combination of disk and planet parameters, the outward (positive)  one-sided horseshoe torque is larger in magnitude than the inward (negative) one-sided Lindblad torque ($\Gamma_{\rm 1s,hs} +   \Gamma_{\rm 1s,L}  >0$).  Since the outer planet is far away from the disk edge in terms of horseshoe units, $(r_{\rm out} -r_{\rm c}) /x_{\rm hs} \gg 1$, its torque is still in the 2-sided regime and negative.  However, the sum of the torques on both planets is dominated by the one-sided torque from the inner planet and evaluates to a net positive value.

Whether or not the planet pair migrates outward is determined by  two timescales: the planet migration timescale $\tau_{\rm m}$ ($= (\dot a/a)^{-1}$, see \eq{adot}) and the cavity expansion timescale $\tau_{\rm c}$ (which approximately equals $3.5 \taud $, see \eq{r_c}).   If the cavity expands at a speed much larger than the migration of the planet, it overtakes the planet.  The planet would then fall into the cavity and cease  migrating. Otherwise, when $\taum<\tauc$, it keeps migrating with the expansion of the cavity.  Because  $\taum < \tauc$ initially,  the total disk torques  drive the outward migration of both planets. However,as $\dot{M}_g(t)$ decreases with time, the disk gas mass diminishes, which increases the migration time $\taum$. Therefore, at some point $\taum(t)$ becomes longer than $\tauc$, which indicates that the disk can no longer accommodate the migration of both planets.  The inner planet therefore drops into the inner cavity radius $r_{\rm c}$  (blue dashed line in \fg{fig2}). We find that this occurs at  $t_{\rm 1}= 3.0 \times 10^{5} \yr$. 

After $t=t_1$ the planet pair initially moves inward, because the net torque is now due to the outer planet and the planets are still in resonance. However, 
the planet pair decouples from MMR when the expanding cavity  approaches the outer planet ($r_{\rm out} \simeq r_{c}$). 
 After that, only the outer planet is able to migrate outward  with the  cavity expansion.  This is because divergent migration leads to resonance crossing rather than resonance trapping \citep{Henrard1983,Murray1999}. Eventually, it ceases migrating at around $5.6 \times 10^{5}~\rm yr $ when the migration timescale of the outer planet becomes longer than the cavity expansion timescale. We denote this time $t_2$. After $t=t_2$ both planets are within the magnetospheric cavity.  

After the gas is entirely depleted,  the planets end up at  $0.116$ AU and $0.254$ AU. Their final  outer-to-inner period ratio ($P_{\rm out}/ P_{\rm in}$)  is $3.24$,  much larger than the original  $3$:$2$ resonance.

\subsection{System  with more massive inner planet} 
We show the result of run 2 in \fg{fig2}{b}. It has identical model parameters as the  default run, but the mass order of the planets is changed: $M_{\rm in} =  6M_{\oplus}$  and  $M_{\rm out} =  3M_{\oplus}$.

Similar to run 1 initially both planets migrate to the cavity edge in the $3$:$2$ MMR.  Also, the  migration timescale for the pair ($\tau_{m}$) is likewise shorter than $\tau_{\rm c}$, so both planets migrate outward  with the expansion of the cavity.    
After a while, the inner planet  ceases migrating when the expansion  rate of the cavity starts to exceed the migration rate of the two resonant planets. This occurs  at  $ t_{1} = 4.6 \times 10^{5}$  yr.  It is later than the time  in run 1 ($ 3.0 \times 10^{5}$ yrs).  When the cavity radius $r_c$ approaches the outer planet at $t_2= 5.7 \times 10^{5}$ yr, it directly falls in, without substantial outward migration as run 1.  Consequently, their final period ratio $P_{\rm out }/ P_{\rm in} = 1.57$, is still close to the $3$:$2$ MMR.

 Why are the final period ratios  between run 1 and run 2  so different when only their mass ratio has changed?  The  small outer-to-inner mass ratio in run 2 has two significant effects. First, the torques' contribution from the inner massive planet becomes even more dominant.  It means that the inner planet falls into the cavity at a later time ($t_1$ in run 2 is larger). Therefore, both planets migrate further out when the inner planet is more massive.   A second consequence is that $r_{\rm c}$ only approaches the outer planet at a late time when the disk mass is relatively low.  A low disk mass and a less massive outer planet leads to subsequent slow outward migration. As a consequence, in run 2, the migration timescale of the outer planet is longer than the cavity expansion timescale when the cavity approaches it.   Hence, the outer planet drops into the cavity, and  their orbits still remain close to the $3$:$2$ MMR.

\subsection{Strong stellar magnetic field} 

We  perform run 3 with the same planet masses and disk parameters as the default run,  but a stronger stellar magnetic field of $B_{\ast} = 5 \kG$. The result is presented in \fg{fig2}c. 
For this strong B-field, $r_{\rm c} = 0.13 \AU$ at the beginning. This is because the magnetospheric cavity is  determined by the dominated stellar magnetic torque. As shown in \eq{r_c}, larger $B_\ast$ results in a stronger magnetic torque that truncates the inner disk further out. The embryos' migration behaviours are similar  to the default run.  They enter into the cavity at the same time ($t_1$ and $t_2$)  and stop at the same period ratio as run 1. The reason is that the time when the planets enter the cavity depends on the comparison between planet migration timescale and  the cavity expansion timescale -- both of which are independent of the $B_{\ast}$.    
However, the stronger B-field carries the planets further away, with the final orbits of the two planets ending up at $0.292 \AU$ and $0.637 \AU$.

\subsection{Light disk} 
Finally, we perform a run with same planet masses as run 1 ($M_{\rm in} = 3 { \ \rm M_{\oplus}},  M_{\rm out} = 6 \rm \ M_{\oplus}$) but different disk parameters ($\dot M_{\rm g0} = 5\times 10^{-10} \rm \ M_{\odot} yr^{-1}$, $\tau_{\rm d} =10^{4} \rm \ yr$). Here the initial disk accretion rare ($\dot M_{\rm g0}$) decreases by a factor of $20$ and the disk  depletion time ($\tau_{\rm d}$) reduces by a factor of $10$.   Low $\dot M_{\rm g0}$ indicates that   the initial disk mass is low and  short $\tau_{\rm d} $ means that  the gas is depleted rapidly. The combination  of these  two parameters implies a light disk (small $\Sigma$ or $q_{\rm d}$) during the  evolution timespan simulated here.   In contrast to the massive disk in the default run,  the light disk can not provide sufficient angular momentum exchange to the planets. Therefore,  the planet migration timescale $\tau_{\rm m }$ is always longer than the cavity expansion timescale $\tau_{\rm c}$ in this case.  Both planets result in only minimal change of  their orbits. The final period ratio of the planet pair is $1.72$, very close to its initial value (see \fg{fig2}d).

\section{Modelling Kepler Systems}
\label{sec:obs}
\label{sec:modelling} 

The goal of this section is to  compare the observed super-Earth systems with our proposed \textit{magnetospheric rebound} model.  In order to investigate under which conditions  these systems preferentially form,  we conduct a parameter study, varying $\dot M_{g}$, $\tau_{\rm d}$ and $B_\ast$.

\subsection{Set-up for the parameters survey}
For our pilot study, we consider two super-Earths systems: Kepler $ 170$ and Kepler $180$. Kepler $170$ contains two super-Earth planets near the $2$:$1$ MMR, whereas the period ratio of the planets in  Kepler 180 is $3.03$.  Theoretically, planet migration theory dictates that planets -- when formed in gas-rich disks -- are likely to end up in MMR. However,  the Kepler mission has shown that many super-Earths are at period ratios  near but not exactly at MMR (\textit{e.g.}, 2.2:1; \citealt{Steffen2015}),  or that their period ratios are far from any MMR.  Motivated by the representative  period ratios of these two systems,  we select them as typical cases and explore why the planets end up so differently.  

The planet masses and their periods,  the stellar masses are listed  in \tb{system}. 
The observed systems are selected from the NASA Exoplanet Archive (http://exoplanetarchive.ipac.caltech.edu/), their planet masses are obtained from \cite{Wolfgang2016}'s  mass-radius relationship (their Eq (1)): 
 \begin{equation}
M_p/M_{\oplus} = 2.1 (R_p/R_{\oplus})^{1.5}
 \label{eq:mass-radius}
\end{equation} 
All planets are initialized in  coplanar and circular orbits. The inner planet is located at the edge of the inner cavity, and the outer planet starts beyond their $2$:$1$ MMR. 

\begin{table*}
    \caption{Properites of observed super-Earth system}
      \centering
    \begin{tabular}{ c c c c c c}
        \hline
         Name   & Stellar mass ($M_\odot$)  & $R_{\rm inn}, R_{\rm out}$  ($R_\oplus$) & $M_{\rm inn}, M_{\rm out}$  ($M_\oplus$) & $P_{\rm inn}, P_{\rm out}$ (day) & $P_{\rm out}/P_{\rm inn}$ \\
           \hline
         Kepler 170 & $0.97 $& $2.94,\ \ 2.72$ &$10.59,\ \  9.42 $& $7.93, \ \ 16.67$ & $2.10$  \\
         Kepler 180 & $1.05 $& $1.51,\ \ 2.69$ &$3.90,\ \  9.27 $& $13.82, \ \ 41.89$  & $3.03$ \\
        \hline
    \end{tabular}
    \label{tab:system}
\end{table*}

%\begin{table*}
%    \caption{initial values for adopted parameters and the meaning of symbols in \fg{Kepler}.\ccc{This table seems redundant and can be entirely removed!}}
%    \centering
%    \begin{tabular}{ c  c  c }
%        \hline
%        \hline
%         $\dot{M}_{\rm g0}  \  (\rm M_{\odot} yr^{-1})$   &  [$ 10^{-9} $, $3 \times10^{-9}$, $10^{-8}$,$3\times 10^{-8}$, $  10^{-9} $ ] \\
%          symbols for $\dot{M}_{\rm g0}$  & upper triangle, down triangle, square, cross, circle \\
%       $\tau_{\rm d} \  (\rm yr)$  &   [$10^3$, \ $10^{4}$, \ $10^5$]   \\
%       color for $\tau_{\rm d}$ & purple, green, orange\\
%        $B_\ast  \ (\rm kG)$   & [ 0.3,\ 1, \ 3]  \\
%        size for $B_\ast$ & small, intermediate, large \\
%        \hline
%        \hline
%    \end{tabular}
%    \label{tab:grid}
%\end{table*}

% The initial disks and stellar parameters  ($\dot M_{g}$, $\tau_{\rm d}$ and $B_\ast$) are given in table \ref{tab:grid}.
  The initial gas accretion ($\dot M_{g}$) rate is sampled by $5$ values in the range of $10^{-9}$ to $10^{-7}  \rm \ M_{\odot} yr^{-1}$, the disk dispersal time ($\tau_{\rm d}$) is given by  $3$ values, $10^3$, $10^4$, $10^5 \rm \ yr$, and the magnetic field strength ($B_\ast$) is selected from 3 values, $0.3$, $1$, $3 \ \rm kG$.  We run $45$ simulations for each system with each simulation referring to one specific combination values of $\dot M_{\rm g0}$, $\tau_{\rm d}$ and $B_\ast$. Their outcomes ($P_{\rm out}/P_{\rm inn}$ and $P_{\rm inn}$) are given  in   \fg{Kepler}.  The  left and right panel show  the results of  Kepler $170$  Kepler $180$, respectively.  The color, size  and  shape of the data points refer to $\tau_{\rm d}$, $B_{\ast}$ and $\dot M_{\rm g0}$, respectively.  The red star marks the location of the observed systems in this $P_{\rm out}/P_{\rm inn}$- $P_{\rm inn}$ diagram. %We quantify the  deviation between  numerical outcomes to the observational data, $\Delta X = | X_{\rm o}-X_{\rm s} |/X_{\rm o}$.   The best fit simulations are defined as   $\Delta (P_{\rm out}/P_{\rm inn})< 3\%$ and $\Delta P_{\rm inn} < 10\%$, which is illustrated  as dashed lines in \fg{Kepler}.

 \begin{figure*}[t]
    \includegraphics[scale=0.5]{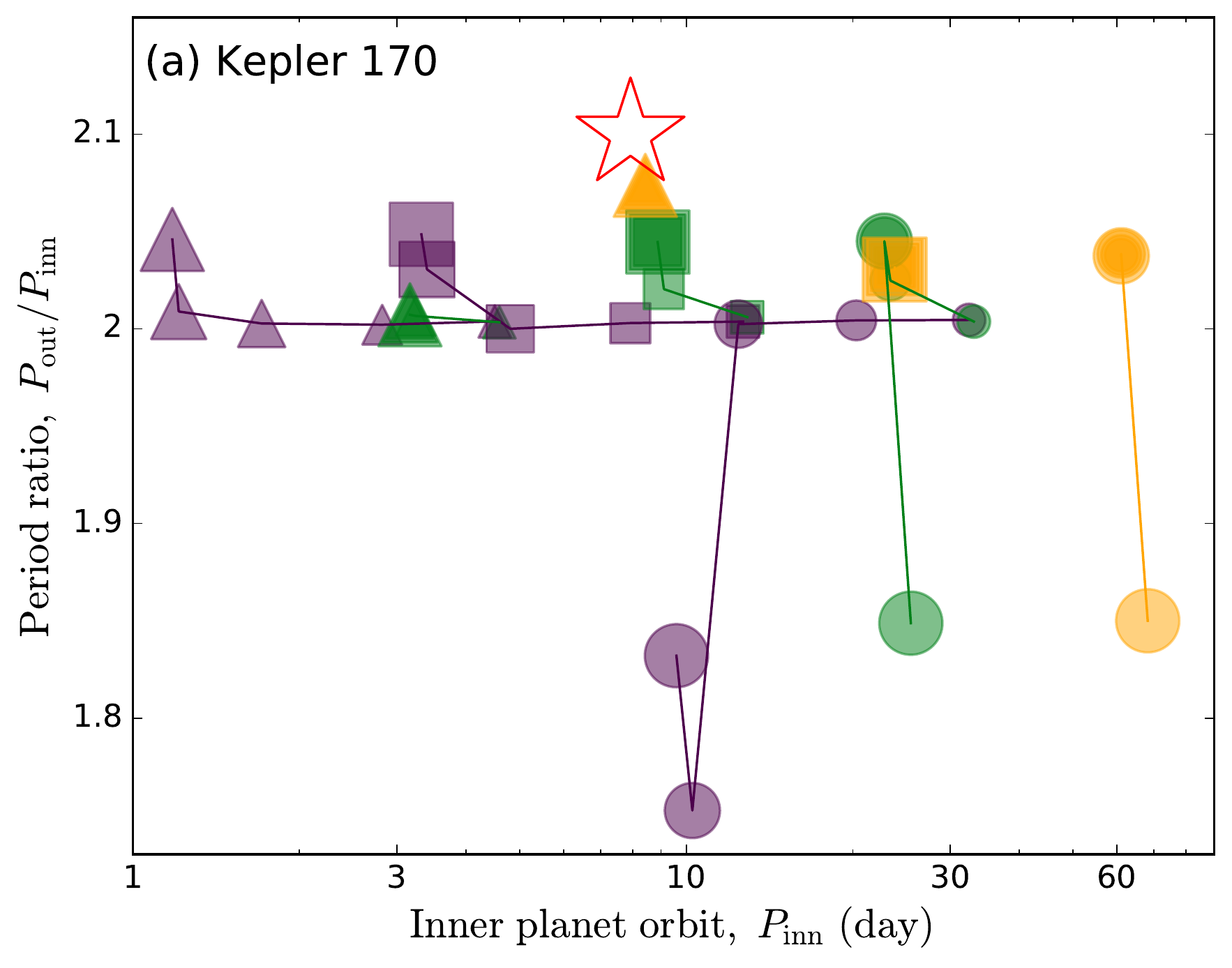}
     \includegraphics[scale=0.5]{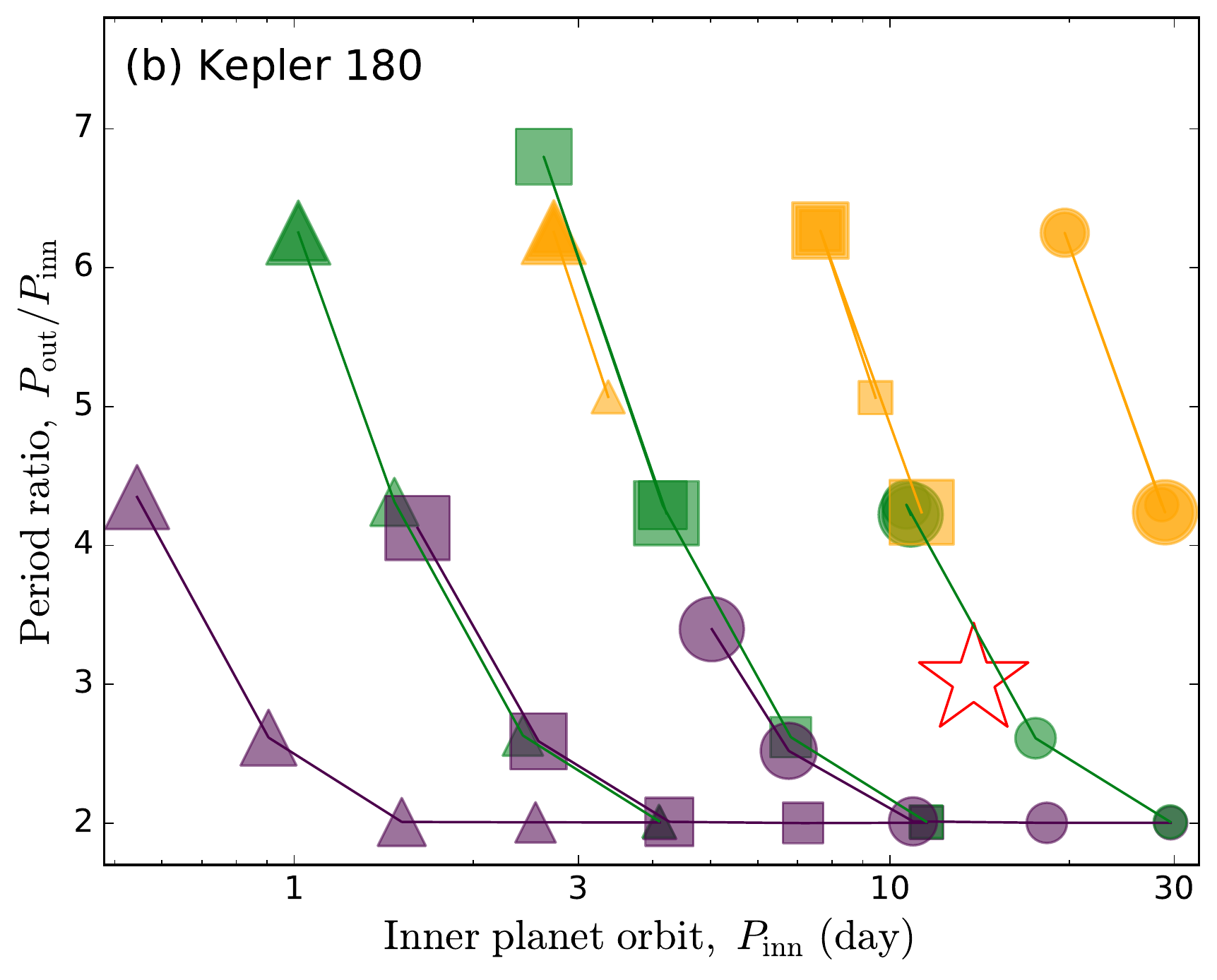}
     \caption{Scatter plot of the inner period of the planet and the period ratio of the outer-to-inner planet.  The left and right panel present  the results of Kepler 170 and Kepler 180.  The red star represents  the observed data and other symbols  give the results of numerical  simulations. Symbols correspond to  stellar magnetic field strengths $B_\ast$ where $B_\ast$ equals $0.3 \kG$ (triangle), $1 \kG$ (square) and  $3 \kG$ (circle). Color corresponds to the disk dispersal timescale ($\tau_{\rm d}$) where purple, green, orange mean $10^{3}$, $10^{4}$ and $10^{5} \yr$.  The size of the symbols represent the disk accretion rate at the onset of disk dispersal ($\dot M_{\rm g0}$), ranging from $ 10^{-9}$ (small),  $3 \times 10^{-9}$, $10^{-8}$, $3 \times 10^{-8}$ to  $10^{-7} \ \rm M_{\odot} yr^{-1}$ (large).
    }
\label{fig:Kepler}
\end{figure*}

\subsection{Kepler 170}
Kepler $170$ is a $0.97 \ M_\odot$  star, containing two super-Earth planets, Kepler $170$b and Kepler $170$c. Based on the radius from the transit and mass-radius relation from \eq{mass-radius}, the masses of the planets are $10.59 \Me$ and $9.42 \Me$, respectively.  Their periods are $7.93$ and $16.67$ days, with a period ratio of $2.10$. 

For this system the simulated data results in a large spread in the period of the inner planet (x-axis of \fg{Kepler}a): $P_\mathrm{inn}$ ranges from $1.2$ to $68$ days. With tuning of the parameters, we can easily obtain a perfect match to the inner planet period. Although the simulated period ratios all lie below the observed value, five of them showing only minor deviation ($\approx$$1.3\%$). Note that all five yellow triangles with different size are stacked at the same period ratio close to $2.07$. One may wonder that the simulations cannot  reproduce  period ratios larger than the observed one. 
The reason is the empirical mass-radius relationship used here (\eq{mass-radius}). Because of the uncertainty in $M(R)$,  the true mass likely deviates from the estimated value. However, a slightly higher outer-to-inner mass ratio-- still consistent with the observational constraints -- will result in a period ratio that matches the observed value.

In \fg{Kepler}a, we see that  for the same initial disk accretion rate and gas depletion timescale (the same size and color), the inner planet period is longer when the stellar B-field is stronger (triangles lie left of circles). This can be understood from our previous discussion related to \fg{fig2}a and c: larger stellar magnetic field strength results in planets stopping their migration at longer periods, because the disk cavity radius $r_c$ is further out. 
 
We also see that for the same gas depletion timescale and stellar B-field but for increasing  $\dot{M}_\mathrm{g0}$  (same symbols from  small to large along the  lines) the period ratio increases from $2$ to $\approx$$2.07$.  Because planets get trapped into the $2$:$1$ MMR first, they migrate outward and leave resonance through the cavity expansion.  A high $\dot{M_{\rm g0}}$  results in substantial outward migration and hence increases their period ratio.  
An exception are the runs with the strongest stellar B-field (circles). Their final period ratios decrease from $2$ to $\approx$$1.8$ with increasing $\dot{M}_\mathrm{g0}$ (large circles in lower right corner of \fg{Kepler}a). The reason is that these planets, \textit{before} the magnetospheric rebound, got trapped in $3$:$2$, instead of $2$:$1$. Because resonance trapping is related to the migration rate, only planets with fast migration rates (meaning: large local disk mass $\Sigma_ p r_p^2$, \eq{adot}) are able to bypass the resonance barrier \citep{Ogihara2013,Liu2015,Liu2016}.  Both a higher disk accretion rate (larger $\Sigma_0$) and  a larger stellar B-field (larger $r_p$)  contribute towards increasing the local disk mass.  Therefore, only systems with  $\dot{M}_\mathrm{g0}$ and large $B_\star$ are able to bypass the $2$:$1$ MMR and be trapped into the $3$:$2$ MMR.  Finally, \textit{magnetospheric rebound} pushes the resonant planets outward and increase the period ratio from the $3$:$2$ MMR to a value between $1.5$ to $2$ during the disk dispersion phase. 
  
The interpretation of this particular system is applicable to systems that contain nearly equal-mass super-Earths,  close to $10$ days inner planet period,  and  period ratio  slightly in excess of 2. These systems follow a similar formation history, and their  $B_\ast$, $\tau_{\rm d}$ and $\dot M_{\rm g0}$ can  be modelled with the same approach shown here.
 
 \subsection{Kepler 180}
 Kepler $180$ is a $1.05 \ M_\odot$ star and harbors two super-Earths: Kepler $180$b and Kepler $180$c. The masses of the planets are $3.90 \Me$ and $9.27 \Me$, and their periods are $13.82 $ and $41.89$ days,  respectively. Their outer-to-inner mass ratio is  $2.38$ and their period ratio is $3.03$. Its result is  presented in \fg{Kepler}b. 

One obvious difference between \fg{Kepler}a and b  is that the period ratios of simulated  planets  in \fg{Kepler}b ($\gtrsim$$2$) can be much larger than those of \fg{Kepler}a ($\sim$$1.7$-$2.1$). 
Although the planets  get quickly trapped into MMR, they move away from resonance during the magnetospheric cavity expansion. We can understand this behavior from the simulations conducted in \fg{fig2}a and \fg{fig2}b (Run \#1,2). There, the only difference is the mass order of the planets.
%, \remove{but} they end up at \change{largely}{very} different period ratios. 
When the outer-to-inner planet mass ratio is large, the outer planet can continue to  migrate  extensively after leaving resonance, resulting in a large period ratio.   
In Kepler $180$ the outer planet is more massive than the inner one, whereas in Kepler $170$ the planets are of nearly equal-mass. Consequently, the period ratios  in \fg{Kepler}b are larger than those in \fg{Kepler}a. 
    
%  Another feature shown in \fg{Kepler}b is that small, purple symbols  are close to the $2$:$1$ MMR, whereas the large, green and orange symbols are far exceed the  $2$:$1$ MMR. It is  because outward migration of planets is substantial in a massive (high $\dot M_{\rm g0}$ and large $\tau_{\rm d}$) disk. Thus, \textit{magnetospheric rebound} mechanism increases   the period ratio of planets in a massive disk. 
%  The period ratios obtained from the light disks (short $\tau_{\rm d}$ and low $\dot M_{\rm g0}$ ) are very close to the $2$:$1$ MMR. On the contrary, the period ratios shown in massive disks (large $\tau_{\rm d}$ and high $\dot M_{\rm g0}$) far exceed the $2$:$1$ MMR.   

From \fg{Kepler}b we see that the observed planet periods and their ratio (red star) lie inside the range spanned up by the simulations.  Therefore, this system is well-fitted by the \textit{magnetospheric rebound} mechanism and we are able to constrain key disk and stellar magnetic parameters.  Kepler 180 represents the  systems that  have a massive outer planet ($M_{\rm out}/M_{\rm inn} \gtrsim 2$) and typical period ratio  larger  than  \noeq{\sim}{2.5}. These super-Earths are far from first order resonance -- a feature that cannot be explained by convergent migration alone.  The \textit{magnetospheric rebound} mechanism that we propose, however, is able to explain the formation of these type of systems.

\section{Discussion and Conclusion}
\label{sec:conclusion}
In this paper,  the \textit{magnetospheric rebound} mechanism has been developed to explain the departure of close-in super-Earths from resonance.  \textit{Magnetospheric rebound} relies on the outward migration of planets that end up near the magnetospheric cavity radius. At this radius (the disk edge) the Type I torque becomes one-sided. We have derived
%First, the expressions of Type I torque formulas provided here  consider  both the cases when the planet is at magnetospheric cavity edge and when the planet is far away from the disk edge.  In particular, 
the expression for the one-sided corotation torque (\eq{hs1}) for the first time.  We have performed N-body simulations of the dynamical evolution of two-planet systems during the gas disk dispersal phase, varying the planet masses ($M_p$), the disk accretion rate ($\dot M_{\rm g0}$), the disk depletion timescale ($\tau_{\rm d}$) and the stellar magnetic field strength ($B_\ast$).  We find that the final orbits of the planets can be substantially rearranged by the magnetospheric cavity expansion.  Three major features of our study are:
\begin{enumerate}
    \item The disk significantly affects the migration behavior and, consequently,  the final period ratio of the planets. The outward migration of the planet with the expanding cavity is substantial in a massive disk  (high $\dot {M_{\rm g}}$, long $\tau_{\rm d}$). On the other hand,  light disks cannot transfer sufficient angular momentum to the planets (low $\dot {M_{\rm g}}$, short $\tau_{\rm d}$). Planets in light disks therefore tend to remain at the orbits they had before the onset of the disk dispersal.
   \item  When the outer planet is more massive than the inner one ($M_{\rm out} >M_{\rm in}$),  the planets tend to move away from their original resonance state. In contrast, when the inner planet is more massive ($M_{\rm out} < M_{\rm in}$), their final period ratio remains similar to the initial one the planets had at the onset of the disk dispersal. 
    \item The strength of the stellar magnetic field ($B_{\ast}$) determines the size of the cavity and consequently affects the planets' final locations. The planets end up at a longer period when the B-field of the central star is stronger.
 \end{enumerate} 

As described in \se{applicability}, our model is limited to the linear approximation of super-Earths (\eq{q_linear}) and the non-gap opening criterion (\eq{q_nogap}). These conditions restrict our model  to low-mass planets ($M_p \lesssim 10M_{\oplus}$ for solar-type stars), hot and  turbulent inner disks ($h \gtrsim 0.025$ and $\alpha_{\nu} \gtrsim 10^{-2}$). Therefore, magnetospheric rebound is inapplicable to  massive  giant planets around solar-type stars, and  super-Earths around very low mass stars.

Although only two-planet systems are considered in this paper, it is straightforward to extend our model  to systems with more than two planets. The timescales analysis (migration timescale of planets \textit{vs.} cavity expansion timescale) given in \se{results} is still applicable to those systems. However, for \noeq{>}{2} planets  dynamical instability may be triggered after the disk is entirely dispersed. Therefore, for \noeq{>}{2} planet systems the long-term (\noeq{\sim}{\mathrm{Gyr}}) secular evolution must be considered. 
In this paper where we consider only two-planet systems, the problem of post-disk dynamical instability does not arise, because our two-planet systems are Hill stable. Specifically, Hill stability is assured when the planets' mutual separation is larger than $2\sqrt{3} R_{\rm H}$ ($P_{\rm out}/P_{\rm inn} \gtrsim 1.06$ for typical super-Earth) \citep{Gladman1993}. This condition is always satisfied for the planets at the end of our simulations.

Overall, our model provides a new direction for studying the dynamics of the super-Earths. For the first time, we have included the effects of the stellar magnetic field, the dispersal of the gas disk and planet migration into one unified model.    Even when  planets  trap into MMRs due to disk migration in the early gas-rich phase, subsequent disk dispersal will induce diversity in their orbital configurations. Therefore, super-Earths can form early and yet escape resonance -- a scenario consistent with both migration theory as well as the Kepler census.

Applying the \textit{magnetospheric rebound} model to Kepler $170$ and Kepler $180$,  we were able to constrain the formation conditions (\textit{e.g.}, $\dot M_{\rm g0}$, $\tau_{\rm d}$ and $B_\ast$) that result in an architecture consistent with the observations (\Se{modelling}).
In addition, our results hint that a two-planet system will end up with a period ratio that depends on the mass order of the planets (\Se{results}). However, to ultimately verify the \textit{magnetospheric rebound} model, a statistical analysis is needed. Specifically, the Kepler mission has vastly increased the number of multi-planetary systems. Therefore, a large sample is available to statistically analyze the orbital properties of the super-Earth population. In a follow-up paper, we will conduct  a full exploration of this mass ratio-period ratio correlation in the light of two formation scenarios of super-Earths \citep{Morbidelli2016}: \textit{in-situ} growth \citep{Hansen2012,Chiang2013} and migration \citep{Terquem2007,McNeil2010}. Comparison between the predictions of the magnetospheric rebound model and the Kepler observations enable us to  distinguish between these formation scenarios, and to gain a better understanding of the origin of super-Earths.

 % This new mechanism would then be applied  to test different formation  scenarios of super-Earths (\textit{e.g.}, \textit{in-situ} formation and migration), and compare the results  with the observed Kepler super-Earth population statistically\ccc{too dense. People can't follow.}.\ccc{this paragraph is a little short. Suggest to merge it with the other short paragraph below \#3} 

\begin{acknowledgements}
We thank  Shigeru Ida, Carsten Dominik, Gijs Mulders, Bin Dai and Zhuoxiao Wang for useful discussions.
We also thank the anonymous referee for  their helpful suggestions and  comments.
 B.L.\ and C.W.O\ are supported by the Netherlands Organization for Scientific Research (NWO; VIDI project 639.042.422).  
 %D.N. L. is supported by UC/Lab grant.

\end{acknowledgements}

\bibliographystyle{aa}
\bibliography{reference}

\begin{thebibliography}{89}
\expandafter\ifx\csname natexlab\endcsname\relax\def\natexlab#1{#1}\fi

\bibitem[{{Aarseth}(2003)}]{Aarseth2003}
{Aarseth}, S.~J. 2003, {Gravitational N-Body Simulations}, 430

\bibitem[{{Armitage}(2010)}]{Armitage2010}
{Armitage}, P.~J. 2010, {Astrophysics of Planet Formation}, 294

\bibitem[{{Armitage}(2011)}]{Armitage2011}
{Armitage}, P.~J. 2011, \araa, 49, 195

\bibitem[{{Artymowicz}(1993)}]{Artymowicz1993}
{Artymowicz}, P. 1993, \apj, 419, 155

\bibitem[{{Balbus} \& {Hawley}(1998)}]{Balbus1998}
{Balbus}, S.~A. \& {Hawley}, J.~F. 1998, Reviews of Modern Physics, 70, 1

\bibitem[{{Baruteau} {et~al.}(2014){Baruteau}, {Crida}, {Paardekooper},
  {Masset}, {Guilet}, {Bitsch}, {Nelson}, {Kley}, \&
  {Papaloizou}}]{Baruteau2014}
{Baruteau}, C., {Crida}, A., {Paardekooper}, S.-J., {et~al.} 2014, Protostars
  and Planets VI, 667

\bibitem[{{Baruteau} \& {Papaloizou}(2013)}]{Baruteau2013}
{Baruteau}, C. \& {Papaloizou}, J.~C.~B. 2013, \apj, 778, 7

\bibitem[{{Batygin} \& {Adams}(2017)}]{Batygin2017}
{Batygin}, K. \& {Adams}, F.~C. 2017, ArXiv e-prints

\bibitem[{{Batygin} \& {Morbidelli}(2013)}]{Batygin2013}
{Batygin}, K. \& {Morbidelli}, A. 2013, \aj, 145, 1

\bibitem[{{Bitsch} \& {Kley}(2010)}]{Bitsch2010}
{Bitsch}, B. \& {Kley}, W. 2010, \aap, 523, A30

\bibitem[{{Bonfils} {et~al.}(2013){Bonfils}, {Delfosse}, {Udry}, {Forveille},
  {Mayor}, {Perrier}, {Bouchy}, {Gillon}, {Lovis}, {Pepe}, {Queloz}, {Santos},
  {S{\'e}gransan}, \& {Bertaux}}]{Bonfils2013}
{Bonfils}, X., {Delfosse}, X., {Udry}, S., {et~al.} 2013, \aap, 549, A109

\bibitem[{{Chatterjee} \& {Ford}(2015)}]{Chatterjee2015}
{Chatterjee}, S. \& {Ford}, E.~B. 2015, \apj, 803, 33

\bibitem[{{Chiang} \& {Laughlin}(2013)}]{Chiang2013}
{Chiang}, E. \& {Laughlin}, G. 2013, \mnras, 431, 3444

\bibitem[{{Cossou} {et~al.}(2014){Cossou}, {Raymond}, {Hersant}, \&
  {Pierens}}]{Cossou2014}
{Cossou}, C., {Raymond}, S.~N., {Hersant}, F., \& {Pierens}, A. 2014, \aap,
  569, A56

\bibitem[{{Davis} {et~al.}(2010){Davis}, {Stone}, \& {Pessah}}]{Davis2010}
{Davis}, S.~W., {Stone}, J.~M., \& {Pessah}, M.~E. 2010, \apj, 713, 52

\bibitem[{{Delisle} {et~al.}(2015){Delisle}, {Correia}, \&
  {Laskar}}]{Delisle2015}
{Delisle}, J.-B., {Correia}, A.~C.~M., \& {Laskar}, J. 2015, \aap, 579, A128

\bibitem[{{Delisle} {et~al.}(2014){Delisle}, {Laskar}, \&
  {Correia}}]{Delisle2014}
{Delisle}, J.-B., {Laskar}, J., \& {Correia}, A.~C.~M. 2014, \aap, 566, A137

\bibitem[{{Delisle} {et~al.}(2012){Delisle}, {Laskar}, {Correia}, \&
  {Bou{\'e}}}]{Delisle2012}
{Delisle}, J.-B., {Laskar}, J., {Correia}, A.~C.~M., \& {Bou{\'e}}, G. 2012,
  \aap, 546, A71

\bibitem[{{Fabrycky} {et~al.}(2014){Fabrycky}, {Lissauer}, {Ragozzine}, {Rowe},
  {Steffen}, {Agol}, {Barclay}, {Batalha}, {Borucki}, {Ciardi}, {Ford},
  {Gautier}, {Geary}, {Holman}, {Jenkins}, {Li}, {Morehead}, {Morris},
  {Shporer}, {Smith}, {Still}, \& {Van Cleve}}]{Fabrycky2014}
{Fabrycky}, D.~C., {Lissauer}, J.~J., {Ragozzine}, D., {et~al.} 2014, \apj,
  790, 146

\bibitem[{{Fang} \& {Margot}(2012)}]{Fang2012}
{Fang}, J. \& {Margot}, J.-L. 2012, \apj, 761, 92

\bibitem[{{Fendyke} \& {Nelson}(2014)}]{Fendyke2014}
{Fendyke}, S.~M. \& {Nelson}, R.~P. 2014, \mnras, 437, 96

\bibitem[{{Frank} {et~al.}(1992){Frank}, {King}, \& {Raine}}]{Frank1992}
{Frank}, J., {King}, A., \& {Raine}, D. 1992, {Accretion power in
  astrophysics.}

\bibitem[{{Fressin} {et~al.}(2013){Fressin}, {Torres}, {Charbonneau}, {Bryson},
  {Christiansen}, {Dressing}, {Jenkins}, {Walkowicz}, \&
  {Batalha}}]{Fressin2013}
{Fressin}, F., {Torres}, G., {Charbonneau}, D., {et~al.} 2013, \apj, 766, 81

\bibitem[{{Ghosh} \& {Lamb}(1979)}]{Ghosh1979a}
{Ghosh}, P. \& {Lamb}, F.~K. 1979, \apj, 232, 259

\bibitem[{{Gladman}(1993)}]{Gladman1993}
{Gladman}, B. 1993, \icarus, 106, 247

\bibitem[{{Goldreich} \& {Schlichting}(2014)}]{Goldreich2014}
{Goldreich}, P. \& {Schlichting}, H.~E. 2014, \aj, 147, 32

\bibitem[{{Goldreich} \& {Tremaine}(1979)}]{Goldreich1979}
{Goldreich}, P. \& {Tremaine}, S. 1979, \apj, 233, 857

\bibitem[{{Goldreich} \& {Tremaine}(1980)}]{Goldreich1980}
{Goldreich}, P. \& {Tremaine}, S. 1980, \apj, 241, 425

\bibitem[{{Hansen} \& {Murray}(2012)}]{Hansen2012}
{Hansen}, B.~M.~S. \& {Murray}, N. 2012, \apj, 751, 158

\bibitem[{{Hartmann} {et~al.}(1998){Hartmann}, {Calvet}, {Gullbring}, \&
  {D'Alessio}}]{Hartmann1998}
{Hartmann}, L., {Calvet}, N., {Gullbring}, E., \& {D'Alessio}, P. 1998, \apj,
  495, 385

\bibitem[{{Henrard} \& {Lemaitre}(1983)}]{Henrard1983}
{Henrard}, J. \& {Lemaitre}, A. 1983, \icarus, 55, 482

\bibitem[{{Howard} {et~al.}(2012){Howard}, {Marcy}, {Bryson}, {Jenkins},
  {Rowe}, {Batalha}, {Borucki}, {Koch}, {Dunham}, {Gautier}, {Van Cleve},
  {Cochran}, {Latham}, {Lissauer}, {Torres}, {Brown}, {Gilliland}, {Buchhave},
  {Caldwell}, {Christensen-Dalsgaard}, {Ciardi}, {Fressin}, {Haas}, {Howell},
  {Kjeldsen}, {Seager}, {Rogers}, {Sasselov}, {Steffen}, {Basri},
  {Charbonneau}, {Christiansen}, {Clarke}, {Dupree}, {Fabrycky}, {Fischer},
  {Ford}, {Fortney}, {Tarter}, {Girouard}, {Holman}, {Johnson}, {Klaus},
  {Machalek}, {Moorhead}, {Morehead}, {Ragozzine}, {Tenenbaum}, {Twicken},
  {Quinn}, {Isaacson}, {Shporer}, {Lucas}, {Walkowicz}, {Welsh}, {Boss},
  {Devore}, {Gould}, {Smith}, {Morris}, {Prsa}, {Morton}, {Still}, {Thompson},
  {Mullally}, {Endl}, \& {MacQueen}}]{Howard2012}
{Howard}, A.~W., {Marcy}, G.~W., {Bryson}, S.~T., {et~al.} 2012, \apjs, 201, 15

\bibitem[{{Johansen} {et~al.}(2012){Johansen}, {Davies}, {Church}, \&
  {Holmelin}}]{Johansen2012}
{Johansen}, A., {Davies}, M.~B., {Church}, R.~P., \& {Holmelin}, V. 2012, \apj,
  758, 39

\bibitem[{{Johns-Krull}(2007)}]{Johns-Krull2007}
{Johns-Krull}, C.~M. 2007, \apj, 664, 975

\bibitem[{{Johns-Krull} {et~al.}(2013){Johns-Krull}, {Chen}, {Valenti},
  {Jeffers}, {Piskunov}, {Kochukhov}, {Makaganiuk}, {Stempels}, {Snik},
  {Keller}, \& {Rodenhuis}}]{Johns-Krull2013}
{Johns-Krull}, C.~M., {Chen}, W., {Valenti}, J.~A., {et~al.} 2013, \apj, 765,
  11

\bibitem[{{Kley} \& {Nelson}(2012)}]{Kley2012}
{Kley}, W. \& {Nelson}, R.~P. 2012, \araa, 50, 211

\bibitem[{{Koenigl}(1991)}]{Koenigl1991}
{Koenigl}, A. 1991, \apjl, 370, L39

\bibitem[{{Lai} {et~al.}(2011){Lai}, {Foucart}, \& {Lin}}]{Lai2011}
{Lai}, D., {Foucart}, F., \& {Lin}, D.~N.~C. 2011, \mnras, 412, 2790

\bibitem[{{Lee} {et~al.}(2013){Lee}, {Fabrycky}, \& {Lin}}]{Lee2013}
{Lee}, M.~H., {Fabrycky}, D., \& {Lin}, D.~N.~C. 2013, \apj, 774, 52

\bibitem[{{Lee} \& {Peale}(2002)}]{Lee2002}
{Lee}, M.~H. \& {Peale}, S.~J. 2002, \apj, 567, 596

\bibitem[{{Lin} \& {Papaloizou}(1979)}]{Lin1979}
{Lin}, D.~N.~C. \& {Papaloizou}, J. 1979, \mnras, 186, 799

\bibitem[{{Lin} \& {Papaloizou}(1993)}]{Lin1993}
{Lin}, D.~N.~C. \& {Papaloizou}, J.~C.~B. 1993, in Protostars and Planets III,
  ed. E.~H. {Levy} \& J.~I. {Lunine}, 749--835

\bibitem[{{Lipunov} \& {Shakura}(1980)}]{Lipunov1980}
{Lipunov}, V.~M. \& {Shakura}, N.~I. 1980, Soviet Astronomy Letters, 6, 28

\bibitem[{{Lithwick} \& {Wu}(2012)}]{Lithwick2012}
{Lithwick}, Y. \& {Wu}, Y. 2012, \apjl, 756, L11

\bibitem[{{Lithwick} {et~al.}(2012){Lithwick}, {Xie}, \& {Wu}}]{Lithwick2012b}
{Lithwick}, Y., {Xie}, J., \& {Wu}, Y. 2012, \apj, 761, 122

\bibitem[{{Liu} {et~al.}(2016){Liu}, {Zhang}, \& {Lin}}]{Liu2016}
{Liu}, B., {Zhang}, X., \& {Lin}, D.~N.~C. 2016, \apj, 823, 162

\bibitem[{{Liu} {et~al.}(2015){Liu}, {Zhang}, {Lin}, \& {Aarseth}}]{Liu2015}
{Liu}, B., {Zhang}, X., {Lin}, D.~N.~C., \& {Aarseth}, S.~J. 2015, \apj, 798,
  62

\bibitem[{{Lopez} \& {Fortney}(2014)}]{Lopez2014}
{Lopez}, E.~D. \& {Fortney}, J.~J. 2014, \apj, 792, 1

\bibitem[{{Mamajek}(2009)}]{Mamajek2009}
{Mamajek}, E.~E. 2009, in American Institute of Physics Conference Series, Vol.
  1158, American Institute of Physics Conference Series, ed. T.~{Usuda},
  M.~{Tamura}, \& M.~{Ishii}, 3--10

\bibitem[{{Marcy} {et~al.}(2014){Marcy}, {Isaacson}, {Howard}, {Rowe},
  {Jenkins}, {Bryson}, {Latham}, {Howell}, {Gautier}, {Batalha}, {Rogers},
  {Ciardi}, {Fischer}, {Gilliland}, {Kjeldsen}, {Christensen-Dalsgaard},
  {Huber}, {Chaplin}, {Basu}, {Buchhave}, {Quinn}, {Borucki}, {Koch}, {Hunter},
  {Caldwell}, {Van Cleve}, {Kolbl}, {Weiss}, {Petigura}, {Seager}, {Morton},
  {Johnson}, {Ballard}, {Burke}, {Cochran}, {Endl}, {MacQueen}, {Everett},
  {Lissauer}, {Ford}, {Torres}, {Fressin}, {Brown}, {Steffen}, {Charbonneau},
  {Basri}, {Sasselov}, {Winn}, {Sanchis-Ojeda}, {Christiansen}, {Adams},
  {Henze}, {Dupree}, {Fabrycky}, {Fortney}, {Tarter}, {Holman}, {Tenenbaum},
  {Shporer}, {Lucas}, {Welsh}, {Orosz}, {Bedding}, {Campante}, {Davies},
  {Elsworth}, {Handberg}, {Hekker}, {Karoff}, {Kawaler}, {Lund}, {Lundkvist},
  {Metcalfe}, {Miglio}, {Silva Aguirre}, {Stello}, {White}, {Boss}, {Devore},
  {Gould}, {Prsa}, {Agol}, {Barclay}, {Coughlin}, {Brugamyer}, {Mullally},
  {Quintana}, {Still}, {Thompson}, {Morrison}, {Twicken}, {D{\'e}sert},
  {Carter}, {Crepp}, {H{\'e}brard}, {Santerne}, {Moutou}, {Sobeck}, {Hudgins},
  {Haas}, {Robertson}, {Lillo-Box}, \& {Barrado}}]{Marcy2014}
{Marcy}, G.~W., {Isaacson}, H., {Howard}, A.~W., {et~al.} 2014, \apjs, 210, 20

\bibitem[{{Mayor} {et~al.}(2011){Mayor}, {Marmier}, {Lovis}, {Udry},
  {S{\'e}gransan}, {Pepe}, {Benz}, {Bertaux}, {Bouchy}, {Dumusque}, {Lo Curto},
  {Mordasini}, {Queloz}, \& {Santos}}]{Mayor2011}
{Mayor}, M., {Marmier}, M., {Lovis}, C., {et~al.} 2011, ArXiv e-prints:
  1109.2497

\bibitem[{{McNeil} \& {Nelson}(2010)}]{McNeil2010}
{McNeil}, D.~S. \& {Nelson}, R.~P. 2010, \mnras, 401, 1691

\bibitem[{{McQuillan} {et~al.}(2014){McQuillan}, {Mazeh}, \&
  {Aigrain}}]{McQuillan2014}
{McQuillan}, A., {Mazeh}, T., \& {Aigrain}, S. 2014, \apjs, 211, 24

\bibitem[{{Mills} {et~al.}(2016){Mills}, {Fabrycky}, {Migaszewski}, {Ford},
  {Petigura}, \& {Isaacson}}]{Mills2016}
{Mills}, S.~M., {Fabrycky}, D.~C., {Migaszewski}, C., {et~al.} 2016, \nat, 533,
  509

\bibitem[{{Morbidelli} \& {Raymond}(2016)}]{Morbidelli2016}
{Morbidelli}, A. \& {Raymond}, S.~N. 2016, ArXiv e-prints:1610.07202

\bibitem[{{Mulders} {et~al.}(2015){Mulders}, {Pascucci}, \&
  {Apai}}]{Mulders2015}
{Mulders}, G.~D., {Pascucci}, I., \& {Apai}, D. 2015, \apj, 798, 112

\bibitem[{{Murray} \& {Dermott}(1999)}]{Murray1999}
{Murray}, C.~D. \& {Dermott}, S.~F. 1999, {Solar system dynamics}

\bibitem[{{Ogihara} {et~al.}(2010){Ogihara}, {Duncan}, \& {Ida}}]{Ogihara2010}
{Ogihara}, M., {Duncan}, M.~J., \& {Ida}, S. 2010, \apj, 721, 1184

\bibitem[{{Ogihara} \& {Ida}(2009)}]{Ogihara2009}
{Ogihara}, M. \& {Ida}, S. 2009, \apj, 699, 824

\bibitem[{{Ogihara} \& {Kobayashi}(2013)}]{Ogihara2013}
{Ogihara}, M. \& {Kobayashi}, H. 2013, \apj, 775, 34

\bibitem[{{Ogihara} {et~al.}(2015){Ogihara}, {Morbidelli}, \&
  {Guillot}}]{Ogihara2015}
{Ogihara}, M., {Morbidelli}, A., \& {Guillot}, T. 2015, \aap, 578, A36

\bibitem[{{Ormel}(2013)}]{Ormel2013}
{Ormel}, C.~W. 2013, \mnras, 428, 3526

\bibitem[{{Paardekooper} {et~al.}(2010){Paardekooper}, {Baruteau}, {Crida}, \&
  {Kley}}]{Paardekooper2010}
{Paardekooper}, S.-J., {Baruteau}, C., {Crida}, A., \& {Kley}, W. 2010, \mnras,
  401, 1950

\bibitem[{{Paardekooper} {et~al.}(2011){Paardekooper}, {Baruteau}, \&
  {Kley}}]{Paardekooper2011}
{Paardekooper}, S.-J., {Baruteau}, C., \& {Kley}, W. 2011, \mnras, 410, 293

\bibitem[{{Paardekooper} \&
  {Papaloizou}(2009{\natexlab{a}})}]{Paardekooper2009a}
{Paardekooper}, S.-J. \& {Papaloizou}, J.~C.~B. 2009{\natexlab{a}}, \mnras,
  394, 2297

\bibitem[{{Paardekooper} \&
  {Papaloizou}(2009{\natexlab{b}})}]{Paardekooper2009b}
{Paardekooper}, S.-J. \& {Papaloizou}, J.~C.~B. 2009{\natexlab{b}}, \mnras,
  394, 2297

\bibitem[{{Papaloizou} \& {Szuszkiewicz}(2005)}]{Papaloizou2005}
{Papaloizou}, J.~C.~B. \& {Szuszkiewicz}, E. 2005, \mnras, 363, 153

\bibitem[{{Petigura} {et~al.}(2013){Petigura}, {Howard}, \&
  {Marcy}}]{Petigura2013}
{Petigura}, E.~A., {Howard}, A.~W., \& {Marcy}, G.~W. 2013, Proceedings of the
  National Academy of Science, 110, 19273

\bibitem[{{Petrovich} {et~al.}(2013){Petrovich}, {Malhotra}, \&
  {Tremaine}}]{Petrovich2013}
{Petrovich}, C., {Malhotra}, R., \& {Tremaine}, S. 2013, \apj, 770, 24

\bibitem[{{Pierens} \& {Nelson}(2008)}]{Pierens2008}
{Pierens}, A. \& {Nelson}, R.~P. 2008, \aap, 482, 333

\bibitem[{{Rein}(2012)}]{Rein2012}
{Rein}, H. 2012, \mnras, 427, L21

\bibitem[{{Rogers}(2015)}]{Rogers2015}
{Rogers}, L.~A. 2015, \apj, 801, 41

\bibitem[{{Romanova} {et~al.}(2002){Romanova}, {Ustyugova}, {Koldoba}, \&
  {Lovelace}}]{Romanova2002}
{Romanova}, M.~M., {Ustyugova}, G.~V., {Koldoba}, A.~V., \& {Lovelace},
  R.~V.~E. 2002, \apj, 578, 420

\bibitem[{{Romanova} {et~al.}(2004){Romanova}, {Ustyugova}, {Koldoba}, \&
  {Lovelace}}]{Romanova2004}
{Romanova}, M.~M., {Ustyugova}, G.~V., {Koldoba}, A.~V., \& {Lovelace},
  R.~V.~E. 2004, \apj, 610, 920

\bibitem[{{Romanova} {et~al.}(2003){Romanova}, {Ustyugova}, {Koldoba}, {Wick},
  \& {Lovelace}}]{Romanova2003}
{Romanova}, M.~M., {Ustyugova}, G.~V., {Koldoba}, A.~V., {Wick}, J.~V., \&
  {Lovelace}, R.~V.~E. 2003, \apj, 595, 1009

\bibitem[{{Shabram} {et~al.}(2016){Shabram}, {Demory}, {Cisewski}, {Ford}, \&
  {Rogers}}]{Shabram2016}
{Shabram}, M., {Demory}, B.-O., {Cisewski}, J., {Ford}, E.~B., \& {Rogers}, L.
  2016, \apj, 820, 93

\bibitem[{{Shakura} \& {Sunyaev}(1973)}]{Shakura1973}
{Shakura}, N.~I. \& {Sunyaev}, R.~A. 1973, \aap, 24, 337

\bibitem[{{Steffen} \& {Hwang}(2015)}]{Steffen2015}
{Steffen}, J.~H. \& {Hwang}, J.~A. 2015, \mnras, 448, 1956

\bibitem[{{Suzuki} {et~al.}(2010){Suzuki}, {Muto}, \& {Inutsuka}}]{Suzuki2010}
{Suzuki}, T.~K., {Muto}, T., \& {Inutsuka}, S.-i. 2010, \apj, 718, 1289

\bibitem[{{Tanaka} {et~al.}(2002){Tanaka}, {Takeuchi}, \& {Ward}}]{Tanaka2002}
{Tanaka}, H., {Takeuchi}, T., \& {Ward}, W.~R. 2002, \apj, 565, 1257

\bibitem[{{Terquem} \& {Papaloizou}(2007)}]{Terquem2007}
{Terquem}, C. \& {Papaloizou}, J.~C.~B. 2007, \apj, 654, 1110

\bibitem[{{Ward}(1997)}]{Ward1997}
{Ward}, W.~R. 1997, \icarus, 126, 261

\bibitem[{{Wardle}(2007)}]{Wardle2007}
{Wardle}, M. 2007, \apss, 311, 35

\bibitem[{{Williams} \& {Cieza}(2011)}]{Williams2011}
{Williams}, J.~P. \& {Cieza}, L.~A. 2011, \araa, 49, 67

\bibitem[{{Winn} \& {Fabrycky}(2015)}]{Winn2015}
{Winn}, J.~N. \& {Fabrycky}, D.~C. 2015, \araa, 53, 409

\bibitem[{{Wolfgang} {et~al.}(2016){Wolfgang}, {Rogers}, \&
  {Ford}}]{Wolfgang2016}
{Wolfgang}, A., {Rogers}, L.~A., \& {Ford}, E.~B. 2016, \apj, 825, 19

\bibitem[{{Xie}(2014)}]{Xie2014}
{Xie}, J.-W. 2014, \apj, 786, 153

\bibitem[{{Xie} {et~al.}(2016){Xie}, {Dong}, {Zhu}, {Huber}, {Zheng}, {De Cat},
  {Fu}, {Liu}, {Luo}, {Wu}, {Zhang}, {Zhang}, {Zhou}, {Cao}, {Hou}, {Wang}, \&
  {Zhang}}]{Xie2016}
{Xie}, J.-W., {Dong}, S., {Zhu}, Z., {et~al.} 2016, Proceedings of the National
  Academy of Science, 113, 41

\bibitem[{{Yang} \& {Johns-Krull}(2011)}]{Yang2011}
{Yang}, H. \& {Johns-Krull}, C.~M. 2011, \apj, 729, 83

\end{thebibliography}

\end{CJK*}
\end{document}